\documentclass[fleqn,usenatbib]{mnras}
\usepackage{newtxtext,newtxmath}
\usepackage[T1]{fontenc}
\DeclareRobustCommand{\VAN}[3]{#2}
\let\VANthebibliography\thebibliography
\def\thebibliography{\DeclareRobustCommand{\VAN}[3]{##3}\VANthebibliography}
\usepackage{graphicx}	
\usepackage{amsmath}	
\usepackage{color}
\usepackage{epstopdf}
\usepackage{float}
\usepackage{lineno}
\usepackage{subfigure}
\usepackage{hyperref}
\usepackage{soul}
\usepackage[normalem]{ulem}

\title[Magnetar with precession motion in GRB 220711B]{Signature of a magnetar central engine with precession motion in the X-ray emission of GRB 220711B}

\author[Shan et al.]{Ying-Ze Shan,$^{1,2}$
Xing Yang,$^{1}$
Hou-Jun L\"{u},$^{1}$
\thanks{Corresponding author (LHJ) email: lhj@gxu.edu.cn}
Jared Rice,$^{3}$
Hao-Yu Yuan,$^{2}$
Xue-Zhao Chang,$^{1}$
\newauthor
Zhao Joseph Zhang,$^{4}$
Le Zou,$^{5}$
and En-Wei Liang,$^{1}$
\\
$^{1}$Guangxi Key Laboratory for Relativistic Astrophysics, School of Physical Science and Technology, Guangxi University, Nanning 530004, China\\
$^{2}$Department of Astronomy, School of Physics, Huazhong University of Science and
Technology, Wuhan, 430074, China\\
$^{3}$Department of Mathematics and Physical Science, Southwestern Adventist University, Keene, TX 76059\\
$^{4}$Department of Earth and Space Science, Osaka University, 1-1 Machikaneyama, Toyonaka, Osaka 560-0043, Japan\\
$^{5}$Department of Physics, Xiangtan University, Xiangtan, Hunan 411105, China\\
}

\date{Accepted XXX. Received YYY; in original form ZZZ}
\pubyear{\the\year{2025}}

\begin{document}
%\label{firstpage}
\maketitle

\begin{abstract}
%\linenumbers
The $\gamma$-ray light curve of long-duration GRB 220711B, is characterized by a multi-peaked structure with a duration lasting $\sim$105 seconds. More interestingly, the X-ray afterglow light curve is composed of a plateau emission smoothly connected with a $\sim t^{-2}$ segment overlapping some flares followed by an extremely steep decay. By analysing the light curves of both prompt emission and X-ray afterglow, no high-confidence-level quasi-periodic oscillation (QPO) signals are found in the light curves of the prompt emission (e.g., BAT and GBM), but it is found that a QPO signal at $\sim$ 50 s above 6$\sigma$ confidence level indeed exist in the X-ray afterglow. Here, we propose that a supra-massive magnetar as the central engine of GRB 220711B with precession motion is a good interpretation of the features of the X-ray emission. The initial plateau emission and followed decay segment, as well as the extremely steep-decay segment, are consistent with the physical process of supra-massive magnetar spin-down and then collapse into black hole. Moreover, the QPO signal in the X-ray emission can be explained as an effect of the precession motion of the magnetar. If this is the case, one can derive various magnetar parameters such as the initial period ($P_{{\rm{0}}}$) and surface magnetic field strength ($B_{{\rm{p}}}$) within a pseudo-redshift range of [1.08, 4.27]. By considering beaming corrections with jet opening angle $5^{\circ}$, we find that $P_{{\rm{0}}}$ and $B_{{\rm{p}}}$ lie within the range of [1.87, 6.25] ms and [$1.47\times 10^{16}$, $3.09\times 10^{16}$] G, respectively. The parameter of $B_{{\rm{p}}}$ is slightly larger than that of other typical long-duration GRBs, but $P_{{\rm{0}}}$ fall in a reasonable range.
\end{abstract}


\section{Introduction}\label{1}
%\linenumbers
Phenomenologically, gamma-ray bursts (GRBs) are classified as long- and short-duration with a separation line at the observed duration $T_{90}\sim 2$ s \citep{1993ApJ...413L.101K}. Some long-duration GRBs (LGRBs) associated with supernovae (SNe) suggest that they are likely related to the deaths of massive stars, and the $``$collapsar$"$ model has been widely recognized as the standard scenario for LGRBs \citep{1993ApJ...405..273W,1998ApJ...494L..45P,1998Natur.395..670G,2003ApJ...591L..17S,2004ApJ...609L...5M,2016ApJ...832..108M,2017A&A...605A.107C}. A small fraction of short-duration GRBs (SGRBs) or SGRB with extended emission (EE) that are claimed to be associated with a fainter-than-supernova optical/infrared transient (called kilonova or mergernova) have been expected to observe \citep{1998ApJ...494L..45P,2010AAS...21530307M,2013ApJ...776L..40Y,2013Natur.500..547T,2015NatCo...6.7323Y,2016NatCo...712898J,2017ApJ...837...50G,2021ApJ...912...14Y,2022ApJ...931L..23L}, and this suggests that they are originated from mergers of two compact stellar objects. The first strong evidence of SGRBs originating in neutron star$-$neutron star (NS–NS) mergers was the discovery of gravitational wave (GW) event GW170817, which was associated with a SGRB 170817A and a kilonova event AT2017gfo \citep{2017PhRvL.119p1101A,2017ApJ...848L..14G,2017ApJ...848L..15S,2018NatCo...9..447Z}.

In addition to the massive star collapse origin for LGRBs, two types of GRB central engine models have been discussed in the literature after the massive star collapse for LGRBs (\citealt{2015PhR...561....1K} for a review). One is a hyper-accreting stellar-mass black hole (e.g., \citealt{1999ApJ...518..356P,2001ApJ...557..949N,2013ApJ...765..125L,2017NewAR..79....1L}), and the other is a rapidly spinning, strongly magnetized neutron star (magnetar) \citep{1992Natur.357..472U,1994MNRAS.270..480T,1998PhRvL..81.4301D,2000ApJ...537..810W,2001ApJ...552L..35Z,2008MNRAS.385.1455M,2012MNRAS.419.1537B,2014ApJ...785...74L,2018MNRAS.480.4402L}.

Within the magnetar LGRB central engine scenario, the magnetar's rotational energy can be lost via electromagnetic (EM) radiation (including the prompt emission and GRB afterglow), continuous GW radiation, or both. The predicted evolution of EM luminosity from the magnetar spin-down is quite different for EM versus GW radiation \citep{2016MNRAS.458.1660L,2018MNRAS.480.4402L}. The GW radiation of magnetar is caused by asymmetric mass distribution, and the rotation is typical on very early stages after the birth of the magnetar. Then, the shape of a magnetar quickly becomes spherical due to its rapid rotation. (1) If the dissipation of the magnetar's rotational energy is dominated by GW radiation then the X-ray luminosity should be a nearly-flat segment followed by $t^{-1}$ decay phase, i.e., $L(t) = L_{0}(1+\frac{t}{\tau_{\rm GW}})^{-1}$, where $\tau_{\rm GW}$ is the characteristic spin-down timescale of the magnetar \citep{2016MNRAS.458.1660L,2018MNRAS.480.4402L}. \cite{2020ApJ...898L...6L} found evidence of GW-dominated emission in the X-ray emission of short GRB 200219A. (2) If the dissipation of the magnetar's rotational energy is dominated by EM radiation then the X-ray luminosity usually manifests as a nearly-flat segment followed by a $t^{-2}$ decay phase, i.e., $L(t) = L_{0}(1+\frac{t}{\tau_{EM}})^{-2}$,  where $L_{0}$ and $\tau_{\rm EM}$ are the initial X-ray luminosity and characteristic spin-down timescale of the magnetar, respectively \citep{2001ApJ...552L..35Z,2016MNRAS.458.1660L,2018MNRAS.480.4402L}. Moreover, the X-ray emission of a small fraction of both LGRB and short GRB show an initial plateau emission followed by steeper power-law decay with slope index less than -3. This is usually interpreted as the magnetar collapsing into a black hole \citep{2007ApJ...665..599T,2010MNRAS.402..705L,2013MNRAS.430.1061R,2014ApJ...785...74L}. 

On the other hand, some models predict the possible existence of quasi-periodic oscillations (QPOs) in light curves of GRBs due to magnetar central engine, activity such as the precession of the magnetar due to the evolution of the inclination angle between the misaligned rotational axis \textbf{$\Omega_{R}$} and magnetic axis \textbf{B} when the magnetar is rapidly rotating \citep{2005ApJ...634L.165S,2012ApJ...760...63L,2014MNRAS.441.1879P,2020ApJ...892L..34S}. Previous studies have performed searches for periodic signals in both the prompt emission and afterglows of GRBs \citep{2002ApJ...576..932K,2010AJ....140..224C,2016A&A...589A..98G,2024ApJ...970....6X}. More interestingly, \cite{2017MNRAS.472.1152R} presented the first targeted deceleration search for periodic signal from spinning down of magnetar central engine, and the non-detection of periodic signal in the X-ray plateaus of the two short GRB may be due to either the reprocessing of the magnetar emission or a strong alignment between the magnetar's rotation axis and the line of sight. However, there have not been any convincing results showing a QPO signal during prompt emission. Recently, \cite{2022MNRAS.513L..89Z} found a low-frequency QPO signal in the X-ray afterglow emission of GRB 180620A. \cite{2023Natur.613..253C} claimed to identify a high-frequency QPO signal in two short GRBs by analyzing archival Burst And Transient Source Experiment (BATSE) data, and a high-frequency QPO signal in three short GRBs are also found in Gamma-ray Burst Monitor (GBM) onboard \textit{Fermi Gamma-ray Space Telescope} (hereafter \textit{Fermi}) (\citealt{2009ApJ...702..791M}; \citealt{2025MNRAS.537.2313Y}). Searches for QPO signals in both the prompt emission and afterglow of GRBs has attracted great attention, and play an important role in understanding the physical properties of magnetars. 

In the ideal case, finding evidence for a magnetar central engine and searching for QPO signals in GRBs will help us understand more details of GRB progenitors and the magnetar formation channel. An interesting case is GRB 220711B, which triggered both the Neil Gehrels \textit{Swift} Observatory (hereafter \textit{Swift})/Burst Alert Telescope (BAT; \citealt{2004ApJ...611.1005G}; \citealt{2005SSRv..120..143B}; \citealt{2022GCN.32366....1D}) and \textit{Fermi}/GBM \citep{2022GCN.32369....1L}. The prompt emission is characterised by a multi-peaked structure and the X-ray afterglow light curve is composed of a plateau emission smoothly connected with a $\sim t^{-2}$ segment, which is the signature of a magnetar central engine. On the other hand, there are some flares that overlap with both the plateau emission and $\sim t^{-2}$ segment. The flares show a quasi-periodic feature, and the power spectra of the X-ray afterglow light curve peaks at $\sim$45 s with a high confidence level (see the section 2.4), which is consistent with the prediction of the magnetar's precession. This paper is organized as follows: the details of data analysis and identification of the QPO signal are presented in Section \ref{2}; in Section \ref{3}, we give a physical interpretation and derive physical parameters. Conclusions are drawn in section \ref{4} with further discussion. Throughout the paper, a concordance cosmology with parameters $H_0 = 70$ $\text{km}$ $\text{s}^{-1} $ $\text{Mpc}^{-1}$ and $\Omega_{M} = 0.3$ is adopted.

\section{Data Reduction and Analysis}\label{2}
In this section, we extract the $\gamma-$ray light curves of GRB 220711B observed by both \textit{Swift}/BAT and \textit{Fermi}/GBM, and perform a spectral fitting. We then adopt the fast Fourier transform (FFT, \citealt{1965MathComp..19..297C}) and Weighted Wavelet Z-transform (WWZ, \citealt{1996AJ....112.1709F}) methods to search for a QPO signal in the gamma-ray prompt emission (\textit{Swift}/BAT and \textit{Fermi}/GBM data) of GRB 220711B. Fourier transform identifies periodicity by decomposing signals into frequency components. FFT is an algorithm to compute Fourier transform by breaking down large transformations into smaller ones, and it use the intermediate results again to reduce computational cost, making it a computationally efficient global analysis method. WWZ algorithm utilizes weighted least squares, a method that applies weighting on different time or frequency intervals to increase the resolution of the wavelet transform. The wavelet transform is a mathematical method that employs a series of wavelets, which are functions that can extract localized information from signals across both time and frequency domains. By comparing with FFT, WWZ is useful to examine the time-resolved power spectrum for a given specific time span. Both FFT and WWZ are widely adopted in QPO analyzes for evenly sampled data. For the \textit{Swift}/X-ray Telescope (XRT; \citealt{2005SSRv..120..165B}) data of GRB 220711B, we present the best fit to the light curve with an empirical function. and then adopt the same method used in BAT and GBM to search for a significant QPO signal.

\subsection{\textit{Swift} data Reduction}\label{2.1}
GRB 220711B triggered the BAT at 18:16:28 UT on 2022 July 11 \citep{2022GCN.32366....1D}. We downloaded the BAT data from the \textit{Swift} website\footnote{\url{https://www.swift.ac.uk/burst_analyser/01115766/}} and used the standard {\sc HEASoft} package (version 6.28) to process the BAT data. The light curves in different energy bands and spectra were extracted by running {\sc batbinevt \citep{2008ApJS..175..179S}}. The light curve shows a multi-peaked structure with duration $T_{90}\sim 105$ s in the 15-350 keV energy range (see Figure \ref{figure:1}) with 1 second time-bin. The time-averaged spectrum is best fit by a simple power-law model with spectral index 1.75 $\pm$ 0.08 due to the narrow energy band \citep{2022GCN.32374....1K}. The XRT data are rebinned to the time resolution of 0.256 s and downloaded by using the re-bin tool in \textit{Swift}/XRT GRB light curve Repository\footnote{\url{https://www.swift.ac.uk/xrt_curves/docs.php\#rebin}}. It worth noting that the rebinning is restricted to the XRT Windowed Timing mode (WT), as the Photon Counting mode (PC) merely supports a maximum time resolution of 2.5 s and does not contain the plateau phase in the X-ray afterglow of GRB 220711B that we are interested in.

\subsection{\textit{Fermi}/GBM data Reduction}\label{2.2}
\begin{figure*}
    \vspace{2cm}
	\centering
	\includegraphics[height=11cm,width=11cm]{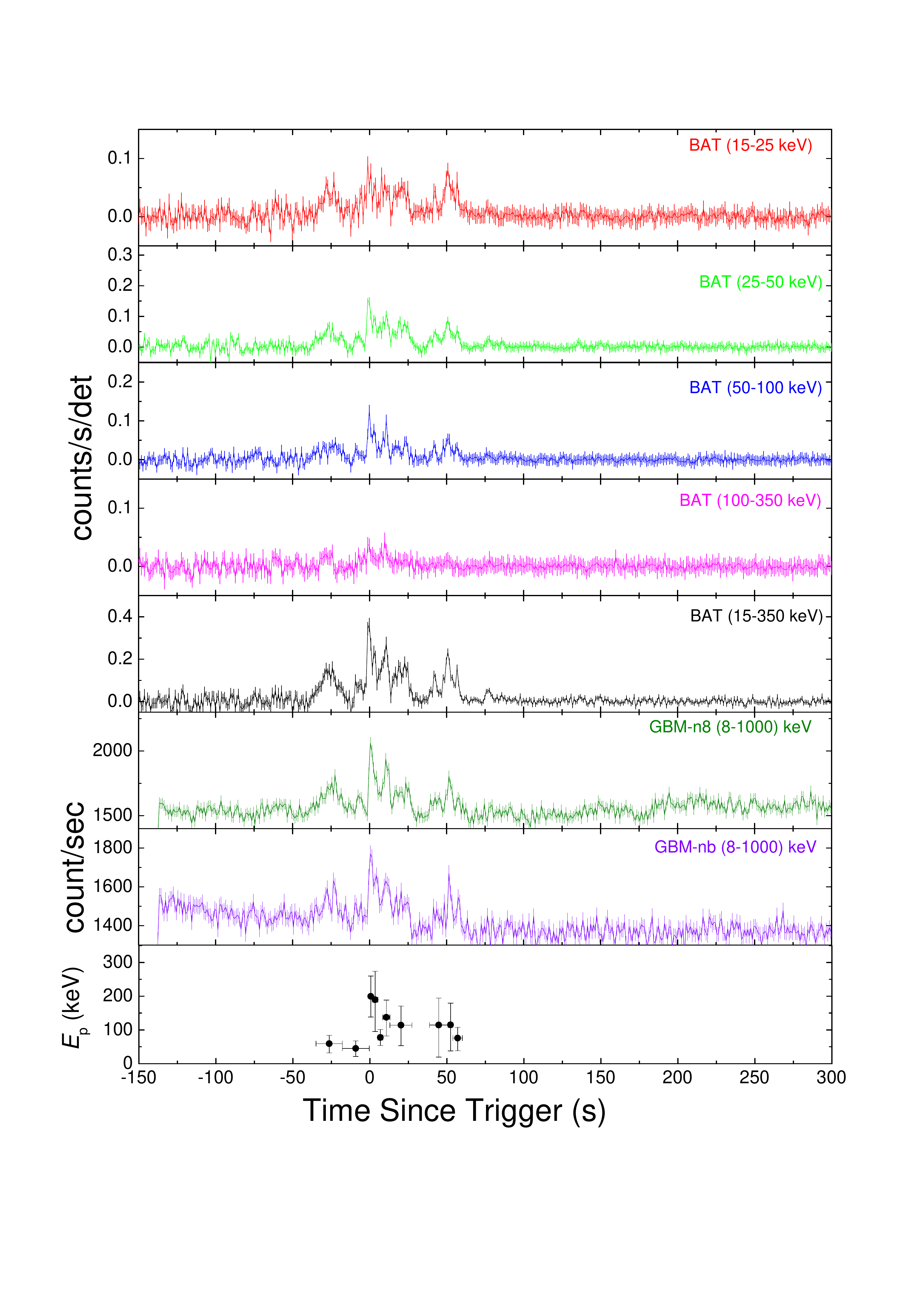}
    \vspace{-10mm}
\caption{\textit{Swift}/BAT and \textit{Fermi}/GBM light curves of GRB 220711B in different energy bands, and the $E_{\rm p}$ evolution.} \label{figure:1}
\end{figure*}

GRB 220711B triggered \textit{Fermi}/GBM at 18:26:45 UT on 2022 July 11 \citep{2022GCN.32369....1L}. We download the corresponding Time-Tagged-Event data from the \textit{Fermi}/GBM public data site\footnote{\url{https://heasarc.gsfc.nasa.gov/FTP/fermi/data/gbm/daily/}}. For more details on the data reduction of the light curve and spectra, refer to \cite{2016ApJ...816...72Z}. The light curves of the n8 and nb detectors are shown in Figure \ref{figure:1}. The light curve shows a multi-peaked structure with duration \textit{$T_{90}\sim 100$} s in the 10-1000 keV energy band.

\begin{figure*}
    \vspace{1cm}
	\centering
	\subfigbottomskip=2pt
	\subfigcapskip=2pt
    \subfigure{{\includegraphics[height=7cm,width=8cm]{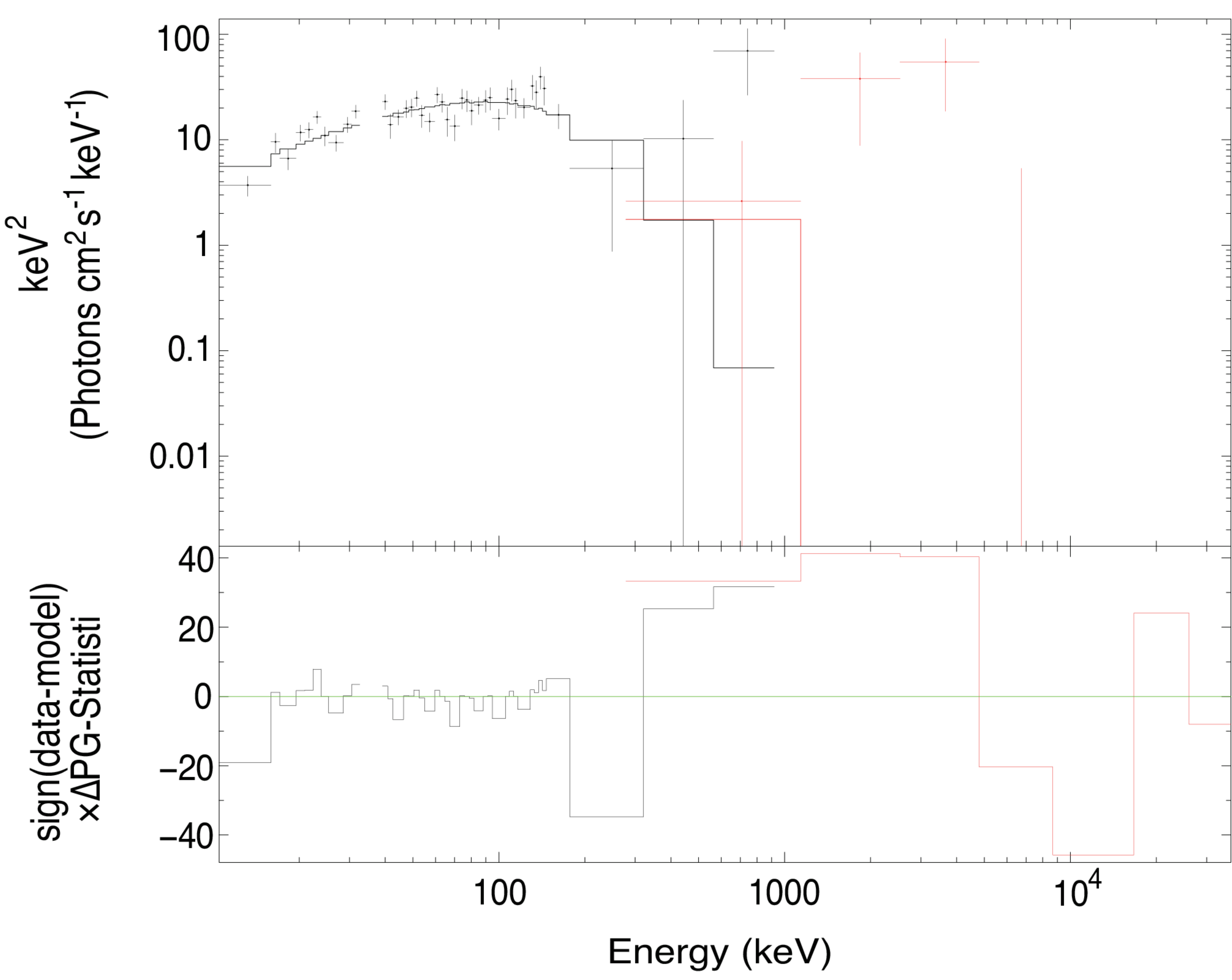}}}
    \hspace{0.5cm}
    \hfill
	\subfigure{\includegraphics[height=7cm,width=8cm]{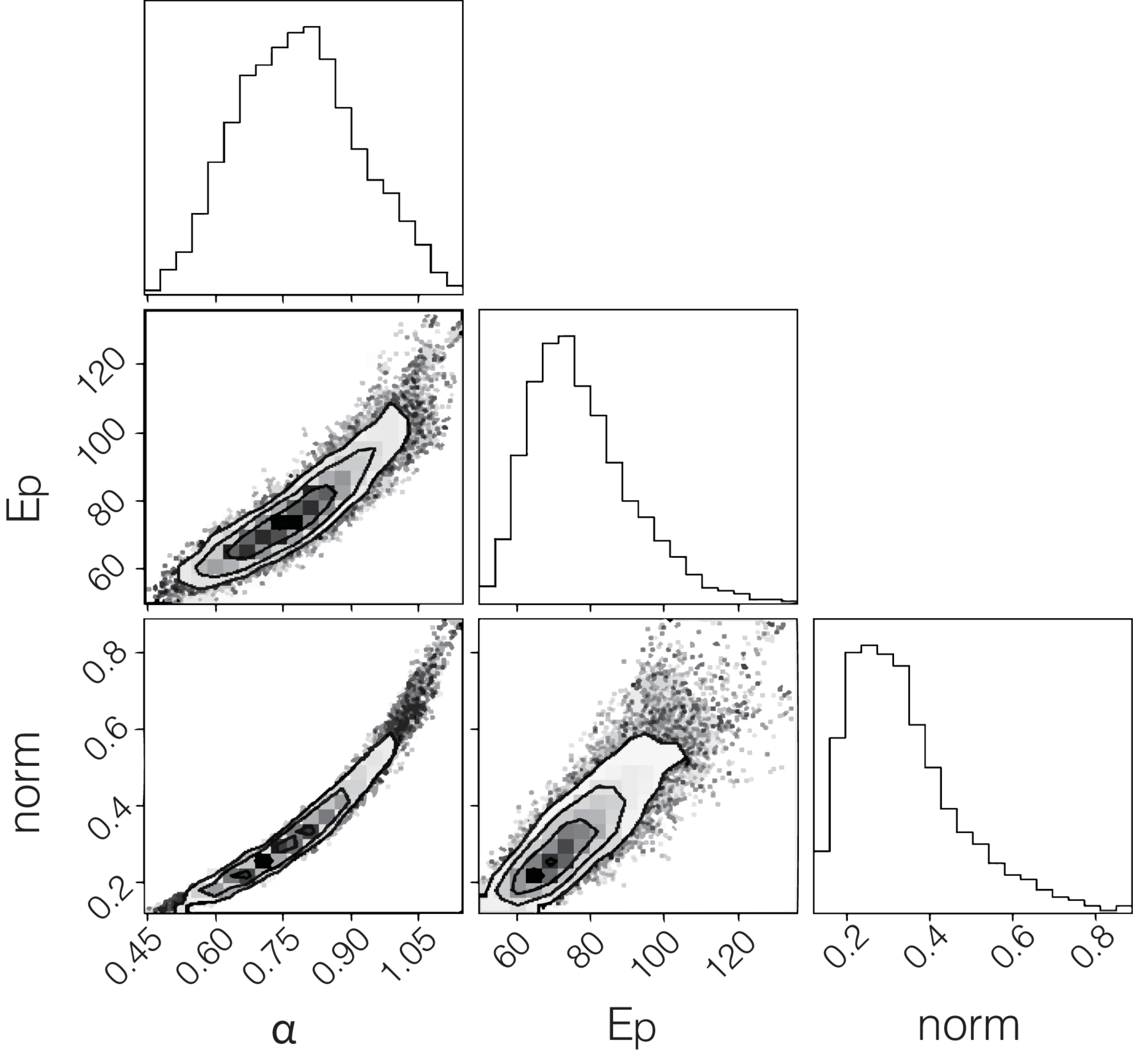}}
\caption{Spectral fits of GRB 220711B with the cut-off power law (CPL) model for \textit{Fermi}/GBM data. The $\nu F_{\nu}$ spectrum and parameter constraints of the CPL fit for the burst are shown in upper and lower panels, respectively. Histograms and contours in the corner plots show the likelihood map of constrained parameters by using the MCMC. The solid black circles from inside to outside are the 1$\sigma$, 2$\sigma$, and 3$\sigma$ uncertainties, respectively.}\label{figure:2}
\end{figure*}

We also extract the time-averaged spectrum of prompt emission (from -40 to 61\,s) in GRB 220711B with the background subtracted \citep{2016ApJ...816...72Z}. We adopt the Markov Chain Monte Carlo (MCMC) method (e.g. \citealt{2013PASP..125..306F}) and multiple empirical models, i.e., power-law function (PL), cutoff power-law function (CPL), Band function (Band), blackbody (BB), and a combination of any of those two to fit the spectral data. We employ the Bayesian information criteria (BIC) to judge the best model among different models \citep{2017ApJ...849...71L}. We compare the BIC and goodness of the fits for different models and find that the CPL model performs best in adequately describing the observed data (see Table \ref{tab:1}). The CPL model can be described as 
\begin{equation}\label{(1)}
N(E,t) = N_{0}(t)\cdot E^{\alpha_{0}}\rm exp(-\frac{E}{E_{\rm p}})
\end{equation}
where $N_{0},$ $E_{\rm{p}}\,$, and $\alpha_{0}\,$ are the normalization of the spectrum, peak energy, and photon index, respectively. The time-integrated spectrum and parameter constraints of the prompt emission CPL model fit is shown in Figure \ref{figure:2}. One has peak energy $E_{\rm{p}}=75\pm 11\,$keV, and a lower energy spectral index of $\alpha_{0}=0.78\pm 0.14$. The best-fit parameters for the CPL fits are listed in Table \ref{tab:1}. Moreover, in Figure \ref{figure:1} we present the time-resolved spectrum of the CPL model, and the $E_{\rm p}$ evolution.

 \subsection{Identification of the QPO in prompt emission}\label{2.3}
Due to the evenly sampled data from BAT and GBM, we adopt the FFT and WWZ methods in the search procedure for QPOs in the prompt emission of GRB 220711B. The n8 detector in \textit{Fermi}/GBM, which has the highest flux among all \textit{Fermi}/GBM detectors, is used to search for QPO signals. We have adopted the {\sc Python 2.7} WWZ package as described in \cite{2017zndo....375648A}. 

\begin{figure*}
    \vspace{2cm}
	\centering
	\subfigbottomskip=20pt
	\subfigcapskip=1.5pt
	\subfigure{\includegraphics[height=6cm,width=8.5cm]{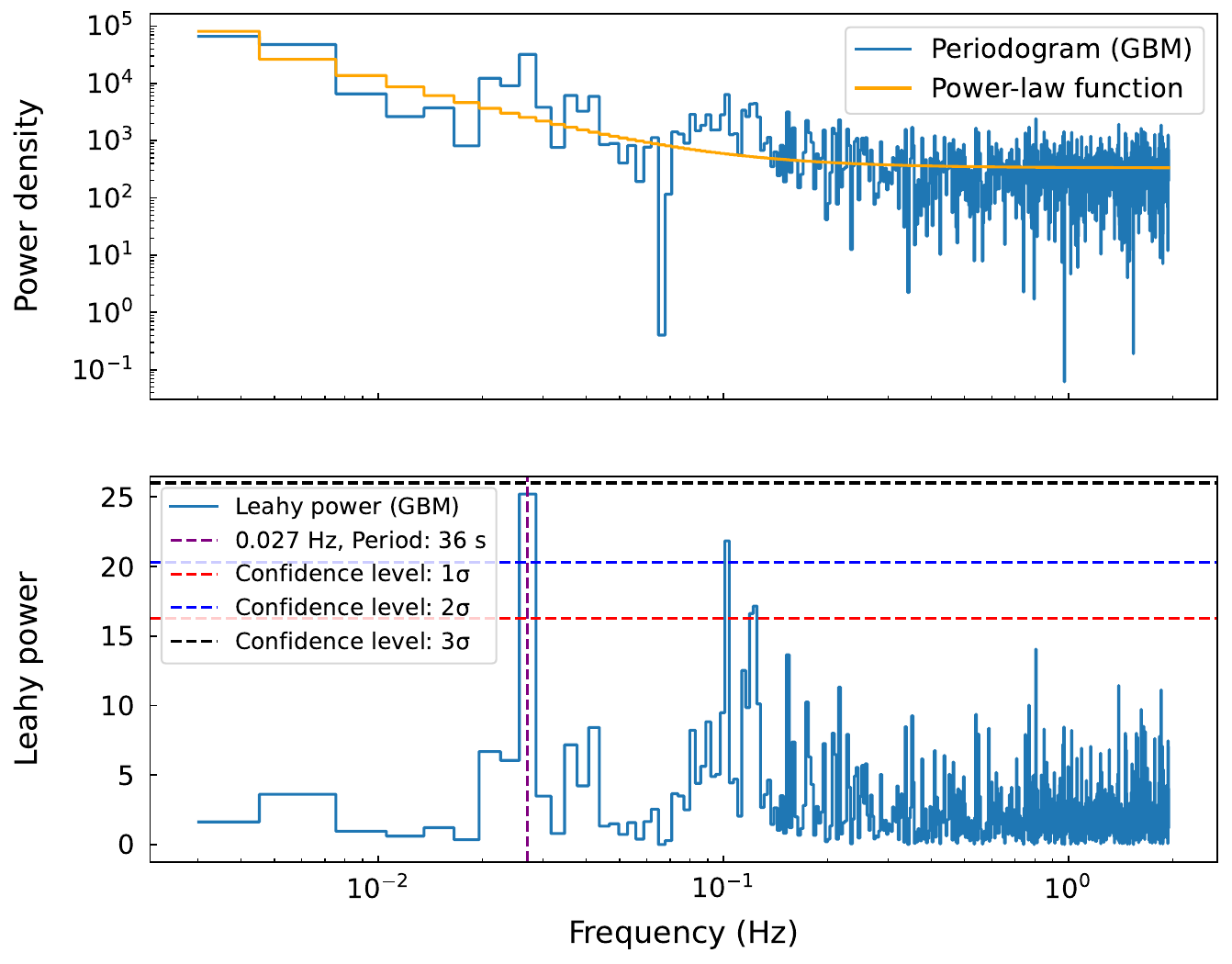}}
	\quad  
	\subfigure{\includegraphics[height=6cm,width=8.5cm]{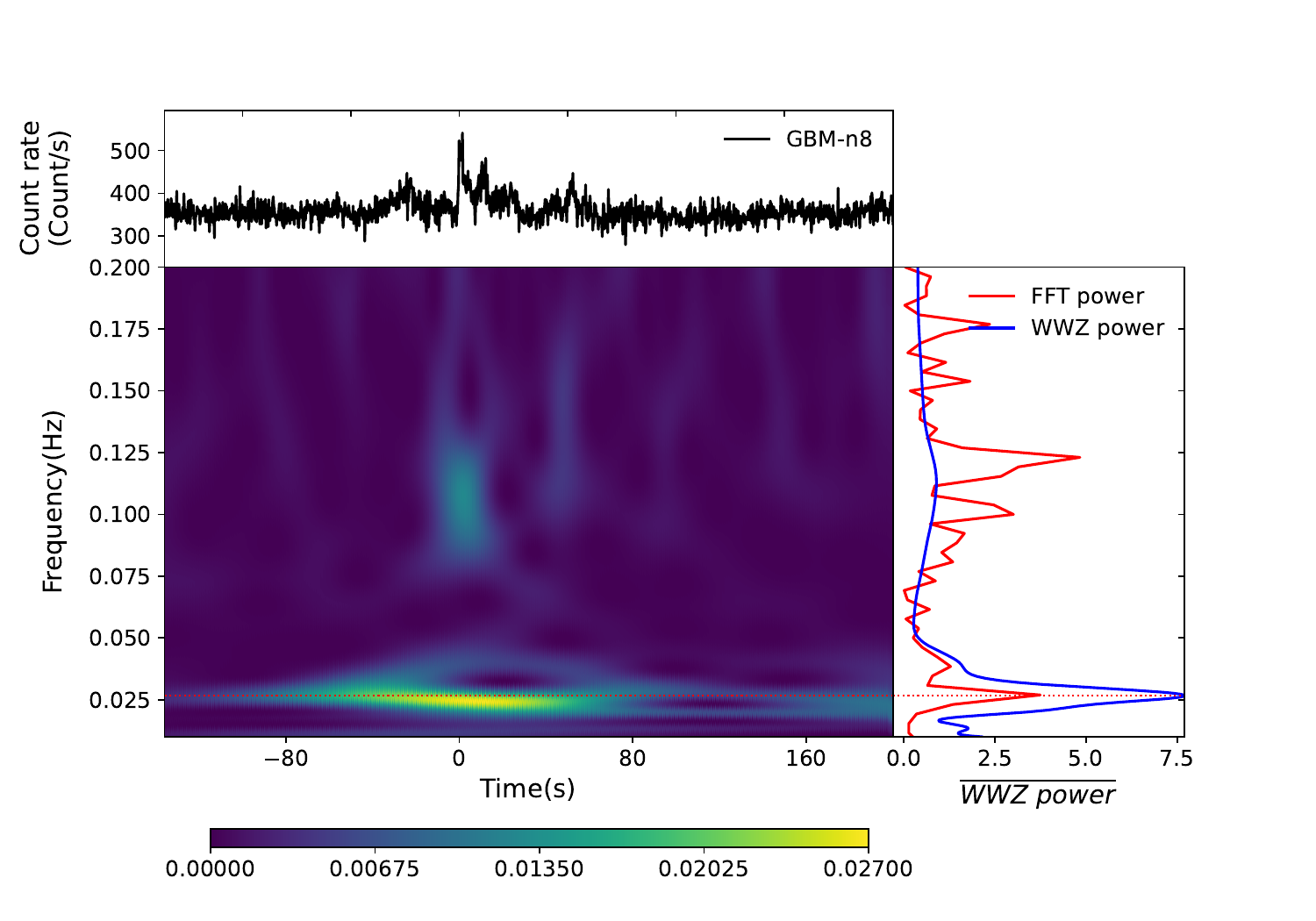}}
    \\
	\subfigure{\includegraphics[height=6cm,width=8.5cm]{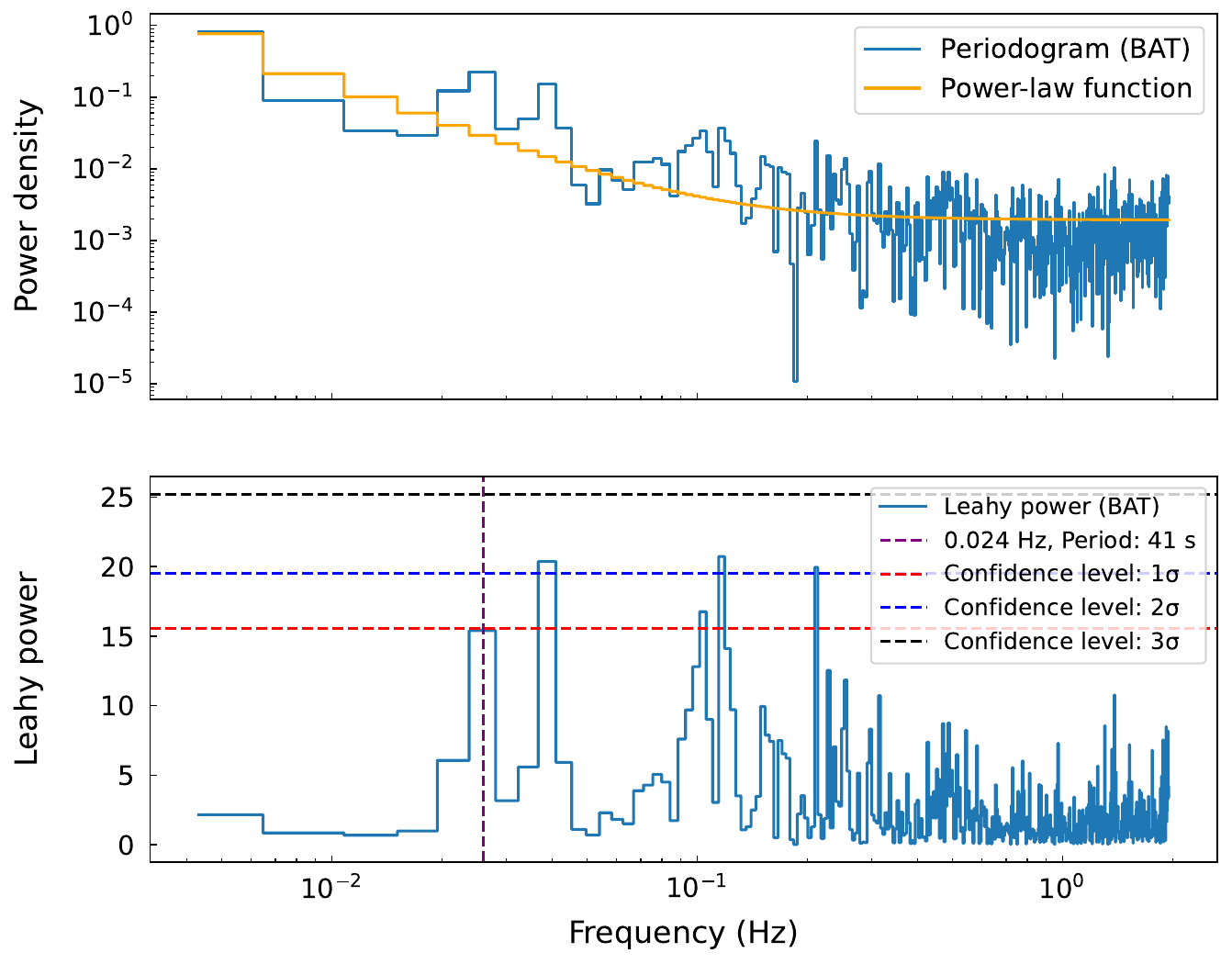}}
	\quad  
	\subfigure{\includegraphics[height=6cm,width=8.5cm]{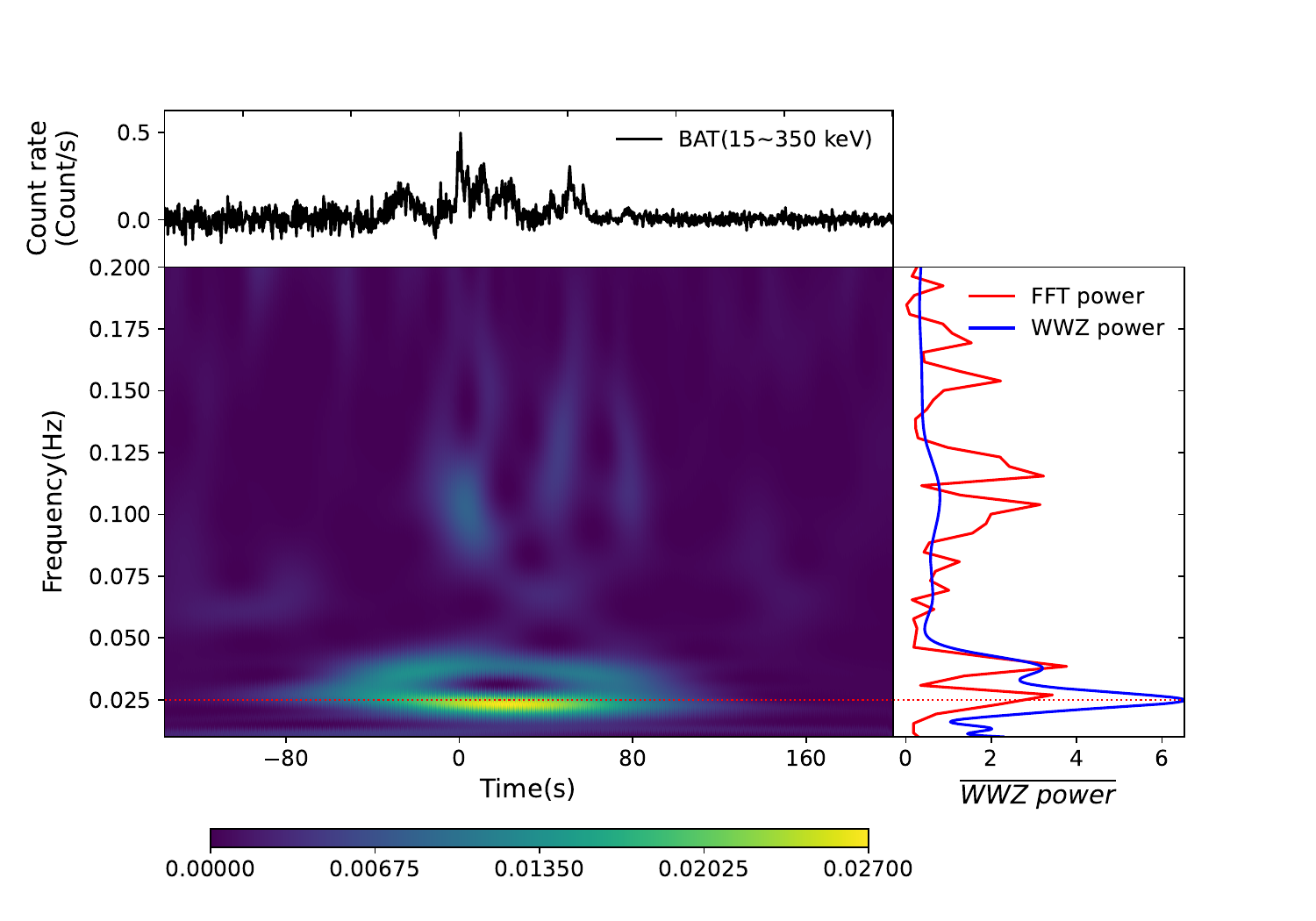}}
     \\
	\subfigure{\includegraphics[height=6cm,width=8.5cm]{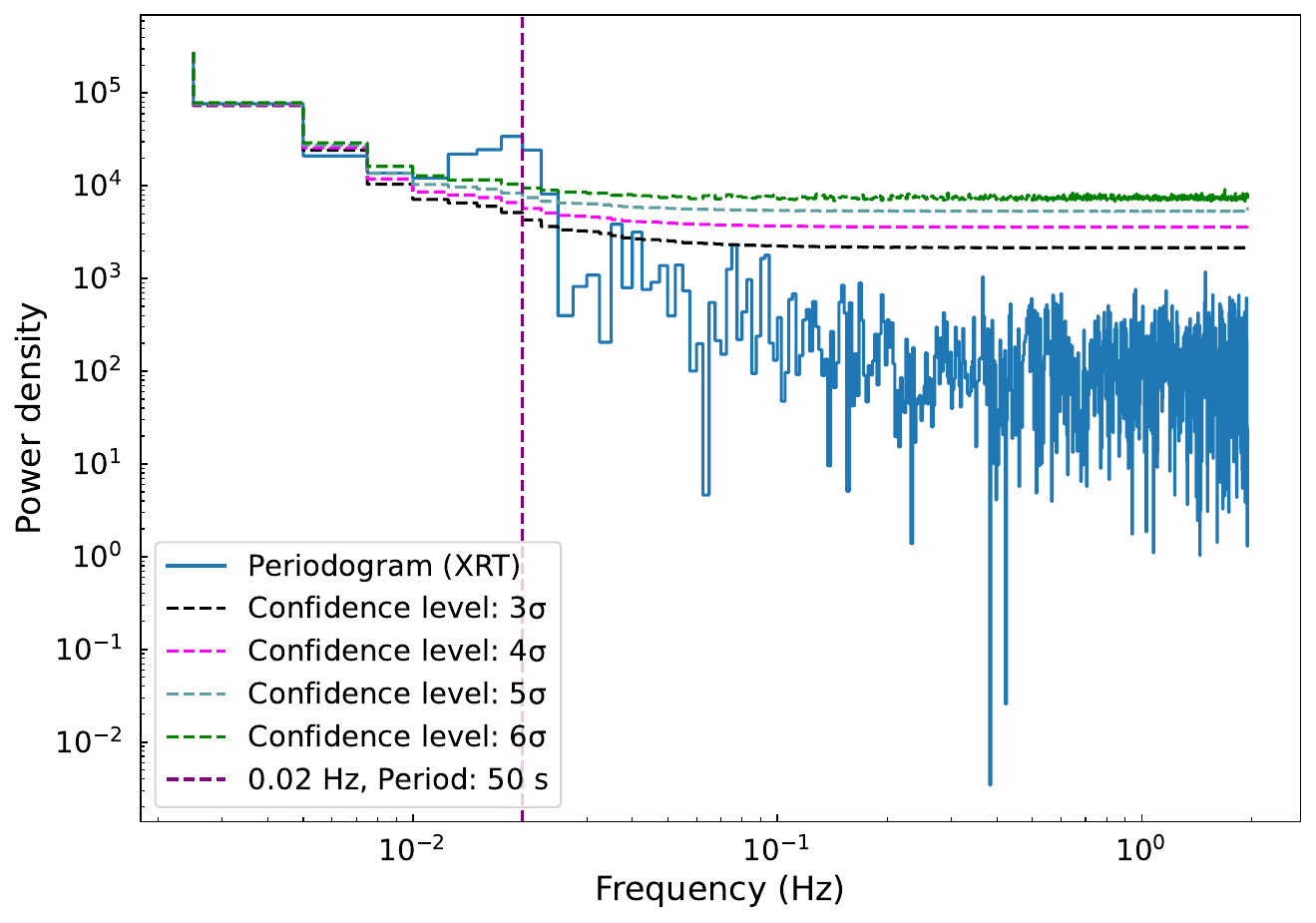}}
	\quad  
	\subfigure{\includegraphics[height=6cm,width=8.5cm]{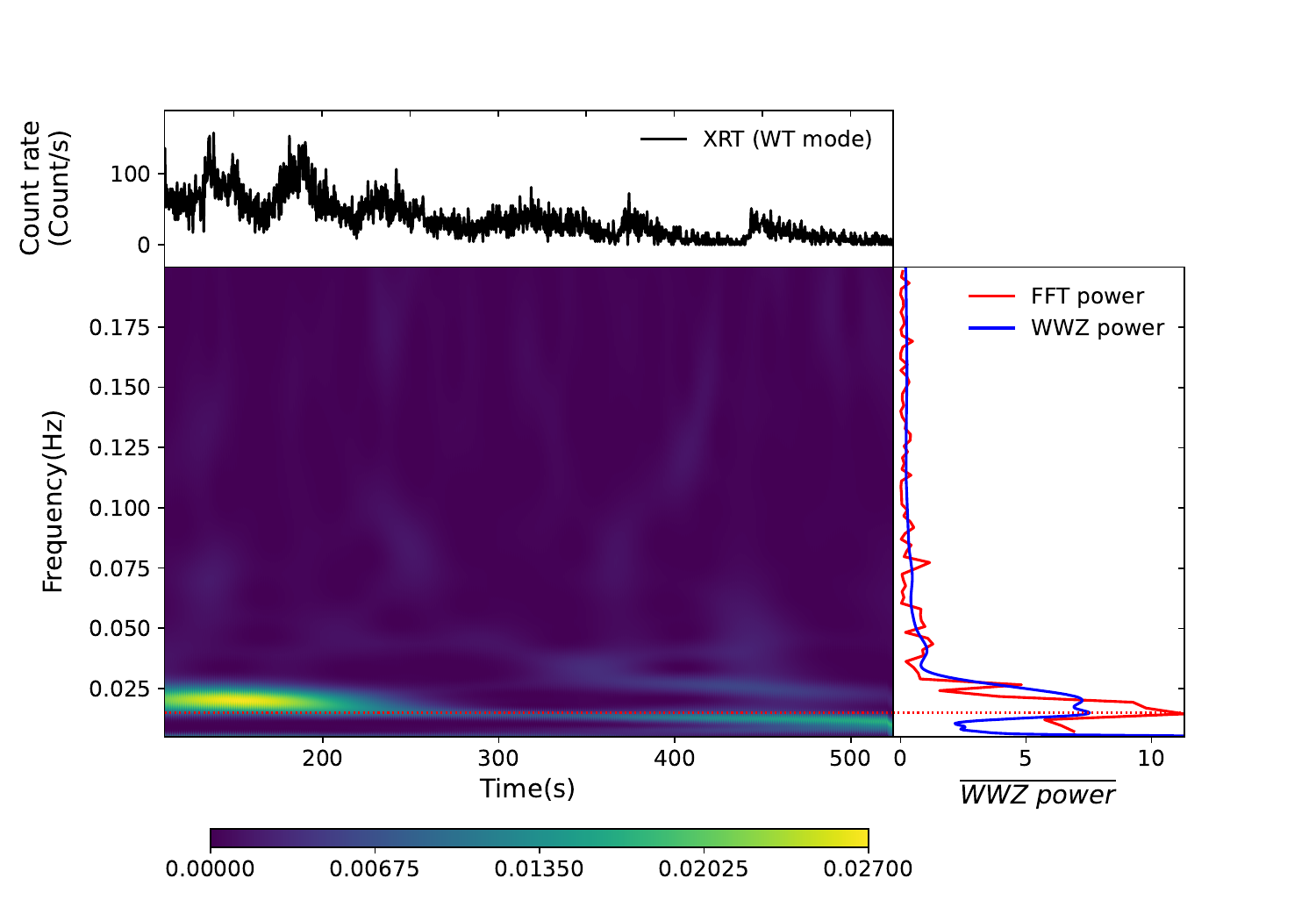}}
\caption{Power spectra of the FFT (left) and time-resolved power spectra of WWZ (right) with QPO signal for the light curve of \textit{Fermi}/GBM (top), \textit{Swift}/BAT (middle), and \textit{Swift}/XRT (bottom). The vertical purple dashed lines in the power spectrum of the FFT and horizontal red dashed lines in the WWZ time-resolved spectrum correspond to the frequencies of peak FFT Leahy power and peak WWZ power, respectively.}\label{figure:3}
\end{figure*}

First, we analyse a potential QPO signal in the n8 detector of the GBM data of GRB 220711B by adopting the FFT method. The periodogram is based on the FFT of the data in an interval of 260 seconds duration, starting 130 seconds before trigger and ending 130 seconds after trigger, with a  0.256 second time bin. The FFT for a time series $x_n$ of $N$ points can be expressed as
\begin{equation}\label{(2)}
a_j=\sum_{n=0}^{N-1} {x_n}e^{2\pi ij \frac{n}{N}}, j=-N/2,...,N/2-1
\end{equation}
The power spectrum is calculated by squaring the modulus of the Fourier transform, also referred to as a periodogram. The power spectrum $I_j$ can be expressed as
\begin{equation}\label{(3)}
I_j={\frac{1}{N}}\sum_{j=0}^{N/2-1} |a_j|^2 , j=1,...,N/2
\end{equation}
The power spectrum of the GBM data can thus be obtained, and normalized to Leahy power (\citealt{1983ApJ...266..160L}, see the top of Figure \ref{figure:3}). The periodogram shows at least two peaks at $\sim$10 seconds and $\sim$36 seconds in the periodogram, respectively.

Then, to confirm the possible periods in the periodogram above, we adopt the WWZ method to examine the time-resolved power spectrum for specific time span. We perform a WWZ analysis on the GBM data within a time range of -136 to 200 seconds and a frequency range of 0.01 to 0.2 Hz. The time and frequency intervals are set to 1 second and 0.0004 Hz, respectively. We find that the bright spot in the WWZ time-resolved power spectra at $\sim$8 seconds (about 0.125 Hz) only lasts for $\sim$20 seconds, which is less than 3 periods. The lack of repeating patterns indicate that the bright spot at 8 seconds is not a real QPO. However, the bright spot at $\sim$40 seconds (about 0.025 Hz) persists throughout the prompt emission (see the middle of Figure \ref{figure:3}), which may correspond to a real QPO. Following the same method as above, we also search for a QPO signal in the \textit{Swift}/BAT data (same time range and binning as GBM) by adopting FFT method, and do not find a significant QPO signal at $\sim$40 seconds. The reason for this may be due to the effect of higher background noise level in the relatively low energy band of BAT (15 keV - 350 keV) when compared to GBM data (10 keV- 900 keV). Similarly, we also adopt the WWZ method to confirm and find that the bright spot at $\sim$40 seconds still exists.

Furthermore, it is necessary to assess the confidence level of the QPO signal in both the GBM and BAT data. Assuming that the photon data follows a Poisson distribution, the ratio between the power spectrum $I_j$ and real spectral spectrum $S_j$ follows an exponential distribution,
\begin{equation}\label{(4)}
f(I_j|S_j)=\frac{1}{S_j}exp(-I_j/S_j)
\end{equation}
The ratio $R^{obs}_j$ between $I_j$ and a parameter-known spectral model $S_j(\theta)$, 
\begin{equation}\label{(5)}
R^{obs}_j=2I^{obs}_j/S_j(\theta)
\end{equation}
follows an identical distribution ($\frac{1}{2}\rm exp(-R^{obs}_j/2)$). It is the same as $\chi^2_n$ distribution with 2 degrees of freedom. The probability density function $f_{\chi^2_n}(x)$ of $\chi^2_n$ distribution is written as
\begin{equation}\label{(6)}
f_{\chi^2_n}(x)=\left\{
\begin{array}{cc}
 \frac{1}{2^{n/2}\Gamma(n/2)}x^{\frac{n}{2}-1} \exp(-x/2),& x>0 \\
 0, & x<=0  
\end{array}
\right.
\end{equation}
By integrating Eq.(\ref{(6)}), the cumulative distribution function $P_j(X<x)$ with 2 degrees of freedom can be expressed as
\begin{equation}\label{(7)}
P_j(Z<z_f)=1-exp(-Z/2)
\end{equation}
$P_j(Z<z_f)$ is the probability that a power caused by a random Poisson noise is higher than that of $Z$ at a single frequency point. By considering all frequency points, one has
\begin{equation}\label{(8)}
F(Z<z_f)=(P_j(Z<z_f))^{N_f}.
\end{equation}
Here, $N_f$ is the number of independent frequencies in the power spectra. The $F(Z<z_f)$ is the probability that a power caused by a random Poisson noise is higher than that of $Z$, i.e. the false alarm probability. We adopt $F(Z<z_f)$ as the confidence level by setting the highest power of prompt emission power spectra as $Z$ (see Figure \ref{figure:3}).

To account for the effect of noise, we adopt a combination of a power-law noise component and a constant background noise component to fit the power spectrum of GRB 220711B prompt emission on a log scale \citep{1995A&A...300..707T,2022ApJ...931...56L}. The combined-noise can be written as $S_{p}(f)=A f^{n}+C$, where $A$ and $n$ are the amplitude and index, respectively. C is the constant representing the white-noise component. It is used to replace $S_{\rm j}$ to recalculate the confidence level. The fitted likelihood function can be written as
\begin{equation}\label{(9)}
L_1(I|\theta,H)=\prod_{j=1}^{N/2} \frac{1}{S_j}\exp{\frac{-I_j}{S_j}}
\end{equation}
This is equivalent to minimizing the following function:
\begin{equation}\label{(10)}
D(I|\theta,H)=-2\ln L_1(I|\theta,H)=2 \sum_{j=1}^{N/2} \frac{I_j}{S_j}+\ln{S_j}
\end{equation}

Finally, we find that the confidence level of the potential QPO signals in the GBM data which peak at $\sim$10 seconds and $\sim$36 seconds are about 3$\sigma$ and 2$\sigma$, respectively (see Figure \ref{figure:3}). For the BAT data, we do not find any significant signals that approach 3$\sigma$. The 2$\sigma$, 3$\sigma$, 4$\sigma$ and 5$\sigma$ correspond to $F(Z<z_f)=$0.95, 0.997, 0.99936 and 0.999999998, respectively. This is only a method to characterize probability, and does not mean that the distribution of powers at each frequency point is Gaussian distribution. Such low confidence level in the potential QPO signals in both the GBM and BAT data is not enough to show convincingly the existence of any QPO in the prompt emission.

\subsection{X-ray observations and fitting}\label{2.4}
\begin{figure*}
    \vspace{1cm}
 	\centering
 	\includegraphics[height=12cm,width=16cm]{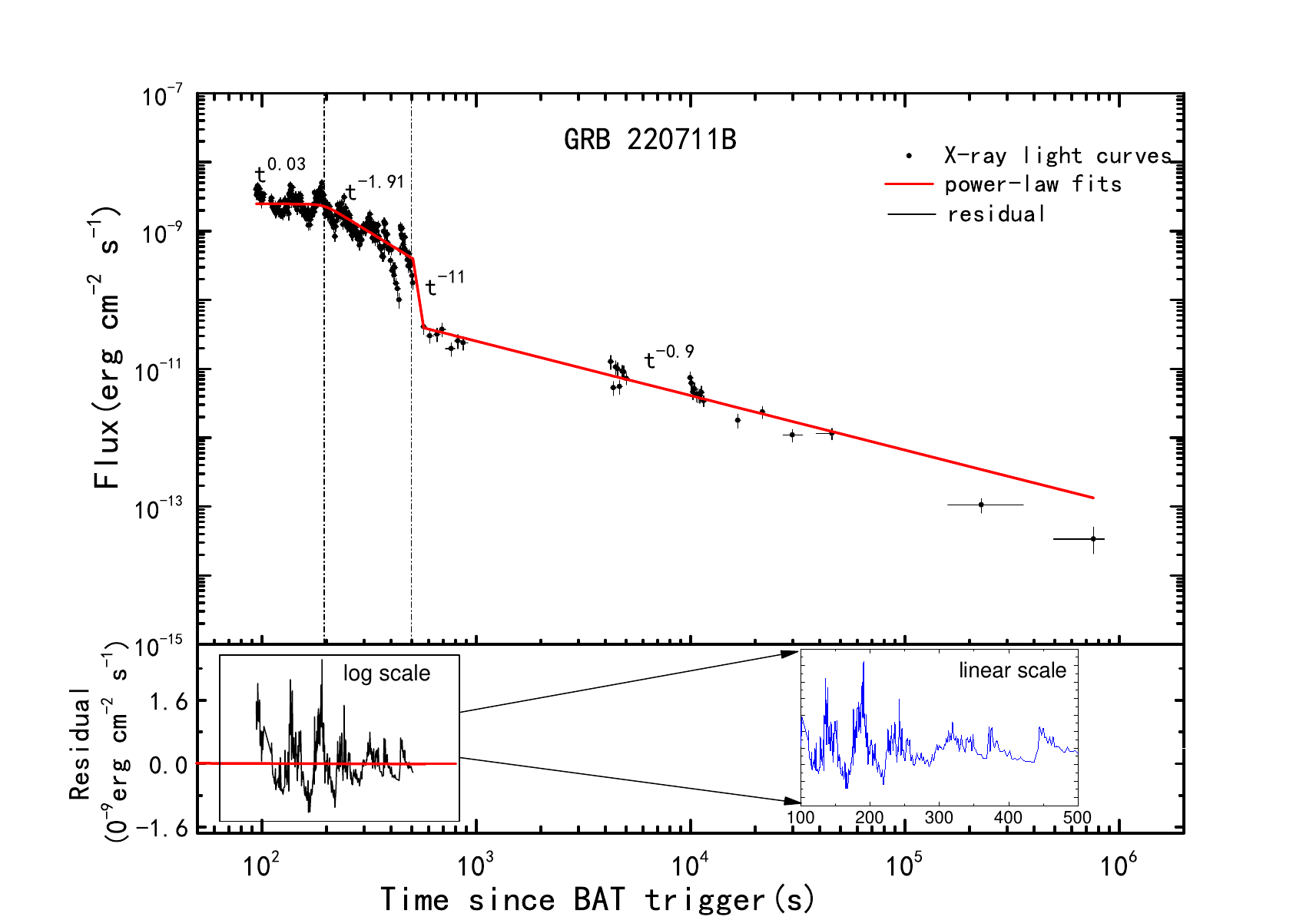}
\caption{X-ray light curve in (0.3-10) keV of GRB 220711B and the best fit with the smooth broken power law functio. Two vertical dashed lines correspond to the breaking time $t_{\rm b}=195\,{\rm{s}}$ and collapse time $t_{\rm col}=517\,{\rm{s}}$} \label{figure:4}
\end{figure*}

\textit{Swift}/XRT began observing the field at 93 s after the BAT trigger \citep{2022GCN.32372....1P}. We made use of public data from the \textit{Swift} archive\footnote{\url{https://www.swift.ac.uk/xrt_curves/01115766/}} \citep{2009MNRAS.397.1177E}. The X-ray light curve is composed of four power-law segments with overlapping flares (see Figure \ref{figure:4}). More interestingly, we find that a QPO signal is likely to be hidden in the flares lasting about 500 seconds after the trigger. In order to confirm the above suspicion, we adopt the empirical fitting to X-ray light curve, and search for the possible QPO signal with the method of WWZ.

First, to dissect the intrinsic features of the early X-ray emission prior to 520 seconds in GRB 220711B, we combined a smoothly broken power-law (SBPL) function with a QPO signal peaking at $45\,{\rm{s}}$ to fit the X-ray light curve. The SBPL function is expressed as,
\begin{equation}\label{(11)}
F_1=F_{0}\left[\left(\frac{t}{t_{\rm b}}\right)^{\omega\alpha_1}+\left(\frac{t}{t_{\rm
b}}\right)^{\omega\alpha_2}\right]^{-1/\omega},
\end{equation}
where $t_{\rm b}$, $\alpha_{1}$, and $\alpha_{2}$ are the break time and slope indices before and after the break, respectively. $\omega$ describes the sharpness of the break and is fixed to 10 \citep{2007ApJ...662.1111L}. We adopted the Sequential Least Squares Programming (SLSQP) which is used an algorithm originally developed by Dieter Kraft to judge our fitting \citep{2020NatMe..17..261V}, and the best fit was found to be $t_{\rm b} = 195$ s, $\alpha_{1} = 0.028$, $\alpha_{2} = 1.91$, and $F_{0}=2.49\times10^{-9}\,\rm{erg/cm^2/s}$ (see Figure \ref{figure:4}). At 500 seconds after the trigger time, the X-ray emission continues to decay with a steeper index $\sim t^{-11}$ followed by a shallow power-law decay with slope index $\sim t^{-0.9}$ (see Figure \ref{figure:4}).

Second, we apply the same QPO searching methods in the Section \ref{2.3} to the X-ray light curve. To search for a higher temporal precision, we perform our analysis using \textit{Swift} X-ray data with a time resolution of 0.256 seconds, which matches the time resolution of GBM and BAT light curves.

Finally, we show the confidence levels for 3$\sigma$, 4$\sigma$, 5$\sigma$, and 6$\sigma$ in the bottom of Figure \ref{figure:3}, and find that the confidence level of the X-ray emission of GRB 220711B is above 6$\sigma$ with a period at $\sim$50 s. One needs to note that the calculated method of confidence level for X-ray is a little different from that of in BAT and GBM. That is because the noise of light curve of BAT and GBM is close to evenly poisson distribution, and the power spectral is power-law function. However, the noise of X-ray emission is unevenly poisson distribution (e.g., a broken power-law function), and the confidence level of QPO in X-ray emission is adopted the simulated light curves to obtain. For example, we simulate the number of $10^{10}$ X-ray light curves (e.g., a broken plower-law function by superposition the poisson distribution), and calculate the power spectral for each light curve. Then, one can obtain the confidence level of 3$\sigma$, 4$\sigma$, 5$\sigma$, and 6$\sigma$. Also, we perform a WWZ analysis on the XRT light curve, and the time range and frequency range are 110$\sim$524 s and 0.005$\sim$0.2 Hz, respectively. Based on the results from the WWZ spectra of XRT light curve (see the bottom of Figure \ref{figure:3}), a bright band at $\sim$50 s (about 0.02 Hz) can be found throughout the time range, which corresponds to the peak power frequency in FFT spectra of XRT light curve. No other bright spots are found in WWZ spectra. 

One question is what is the difference in the red noise between GRB 220711B and other GRB X-ray afterglows detected by \textit{Swift}/XRT? Here, we adopt a combination of a red-noise power-law and a white-noise amplitude to fit the noise of each GRB X-ray afterglow \citep{2022ApJ...931...56L}, and make a statistical distribution of the power-law index for the red noise fit. We find that the power-law index of the red-noise fit for other GRB afterglows lies in the range -3.4 to -1.1, and peaks at -2.2. The power-law index of the red noise for the GRB 220711B X-ray afterglow is -1.95, which lies in the range of this index for other GRB afterglows.

\section{Physical interpretations}\label{3}
\subsection{Pseudo-redshift measured}\label{3.1}
\begin{figure*}
    \vspace{1cm}
	\centering
	\includegraphics[height=9cm,width=15cm]{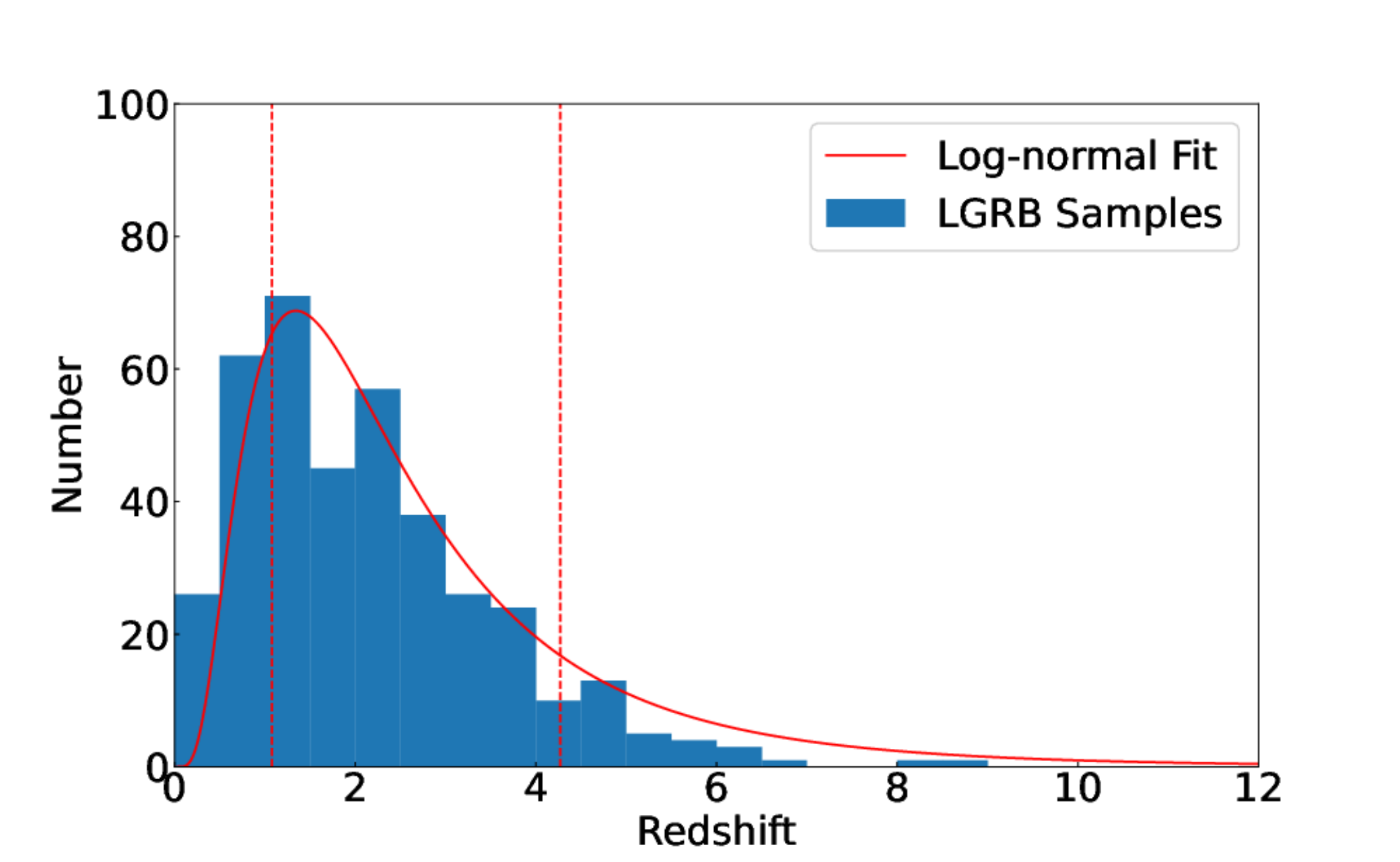}
\caption{The redshift distribution of long GRBs observed by \textit{Swift}. The red curve is the log-normal fit of the redshift distribution. The two red dashed lines show the 1$\sigma$ region ([1.08, 4.27]) for the log-normal distribution by fitting.}\label{figure:5}
\end{figure*}

To understand the intrinsic total energy released by GRB 220711B, measuring the redshift is very important. However, the redshift of GRB 220711B is unknown. Here, we assume that the LGRB 220711B originated in a massive star collapse. Firstly, we collect the redshifts of all LGRBs with redshift observed by \textit{Swift}\footnote{\url{https://swift.gsfc.nasa.gov/archive/grb_table/}}, and get 387 samples in total. Then, we make a distribution of these redshift samples of LGRBs, and adopt a log-normal distribution to fit this distribution. Finally, the $1\sigma$ range of the redshift distribution is calculated from the fitting result, which is from 1.08 to 4.27. The distribution of \textit{Swift} LGRB redshift samples, along with the log-normal fitting result and the $1\sigma$ range are shown in Figure \ref{figure:5}. Therefore we adopt the redshift range $1.08<z<4.27$ to perform the calculations of the following.

\subsection{Magnetar central engine}\label{3.2}
The total energy budget of a newly born magnetar is the rotational energy,
\begin{equation}\label{(12)}
	E_{{\rm {rot}}}=\frac{1}{2}I\Omega^{2}\simeq 2\times 10^{52}\;{{\rm {erg}}}\,M_{1.4}\,R^{2}_{6}\,P^{-2}_{0,-3},
\end{equation}
where $I$, $\Omega$, $P_0$, $R$, and $M$ are the moment of inertia, angular frequency, rotational period, radius, and mass of the magnetar, respectively. The convention $Q_{m}=Q/10^{m}$ is adopted in cgs unit throughout the paper. In general, the magnetar loses its rotational energy via EM emission ($L_{{\rm {EM}}}$) and GW radiation ($L_{{\rm {GM}}}$), which can be expressed as \citep{2001ApJ...552L..35Z,2013PhRvD..88f7304F,2016MNRAS.458.1660L,2018MNRAS.480.4402L} 
\begin{equation}\label{(13)}
\begin{aligned}
	-\frac{{\rm {d}}E_{{\rm {rot}}}}{{\rm {d}}t}&=-I\Omega\dot{\Omega}=L_{{\rm {EM}}}+L_{{\rm {GW}}}\\
		  &=\frac{B_{{\rm {p}}}^{2}R^{6}\Omega^{4}}{6c^{3}}+\frac{32GI^{2}\epsilon^{2}\Omega^{6}}{5c^{5}},
\end{aligned}
\end{equation}
where $c$, $B_{{\rm {p}}}$ and $\epsilon$ are the speed of light, the surface magnetic field and ellipticity of the magnetar, respectively. If the rotational energy loss is dominated by EM radiation, one can easily
derive the luminosity evolution as 
\begin{equation}\label{(14)}
	\begin{aligned}
		L_{{\rm {EM}}}(t)&=L_{{\rm {em,0}}}\left(1+\frac{t}{\tau_{{\rm{c,em}}}}\right)^{-2}\\
	\end{aligned}
\end{equation}\\
where $\tau_{{\rm{c,em}}}$ and $L_{{\rm {em,0}}}$ are the characteristic timescale and initial luminosity of the magnetar spin-down,
\begin{equation}\label{(15)}
	\tau_{{\rm{c,em}}}=\frac{3c^{3}I}{B_{{\rm {p}}}^{2}R^{6}\Omega_{{\rm{0}}}^{2}}=2.05\times 10^{3}{{\rm {s}}}\, I_{{\rm{45}}}\, B_{{\rm {p,15}}}^{-2}\, P_{{\rm{0,-3}}}^{2}\, R_{{\rm{6}}}^{-6},
\end{equation} 
\begin{equation}\label{(16)}
	L_{{\rm {em,0}}}=\frac{I\Omega_{{\rm{0}}}^{2}}{2\tau_{{\rm{c,em}}}}\simeq 1.0\times 10^{49}\, {{\rm {erg\, s^{-1}}}}\, B_{{\rm {p,15}}}^{2}\, P_{{\rm{0,-3}}}^{-4}\, R_{{\rm{6}}}^{6}.
\end{equation}

From the observational point of view in section \ref{2.4}, the early X-ray light curve is composed of an initial plateau emission followed by a steep decay with slope $\sim t^{-2}$. This feature is quite consistent with the prediction of the magnetar central engine rotational energy loss above. If this is the case, one can roughly estimate $P_{{\rm{0}}}$ and $B_{{\rm{p}}}$ for GRB 220711B by adopting the fitting results of the X-ray afterglow from section \ref{2.4}, i.e. $F_0=2.49\times10^{-9}\,\rm{erg/cm^2/s}$ and $t_{\rm b}=195\,\rm{s}$. To calculate $L_{{\rm{em,0}}}$ and $\tau_{{\rm{c,em}}}$ in the rest frame of magnetar, the k-correction \citep{2001AJ....121.2879B}, a method which convert the observed magnitudes of distant objects to their rest-frame values by accounting for the redshift-dependent shift in spectral energy distribution, should be considered. The k-correction factor $k_{\rm cz}$ is calculated by
 \begin{equation}\label{(17)}
	k_{\rm cz}=\frac{\int_{{1}/{(1+z)}}^{{10^4}/{(1+z)}} E N(E) \, dE}{\int_{0.3}^{10} E N(E) \, dE}.
\end{equation} 
The 0.3-10 keV XRT-band is corrected to 1-$10^{4}$ keV in the rest frame. We adopt the {\sc Python} package {\sc REDBACK} \citep{2024MNRAS.531.1203S} to perform the k-correction for the pseudo-redshift (from $z=1.08$ to $z=4.27$), and the photon index obtained from \textit{Swift} Windowed Timing (WT) mode of X-ray light curve is 1.89 in the website\footnote{\url{https://www.swift.ac.uk/xrt_spectra/01115766/}}. Since the observed X-ray luminosity can be calculated by using the parameters of fitting the light curve with a broken power-law,
\begin{equation}\label{(18)}
	L_{\rm X}(t)=4\pi D_{\rm L}^{2}F(t)=\eta_{\rm X}L_{{\rm{em,0}}}(1+\frac{t}{\tau_{\rm{c,em}}})^{-2},%\left(1+\right)
\end{equation} 
the $L_{{\rm{em,0}}}$ and $\tau_{{\rm{c,em}}}$ can thus be derived by the plateau flux and duration from the X-ray lightcurve fitting:
\begin{equation}\label{(19)}
    L_{{\rm {em,0}}}=4\pi D_{\rm L}^{2}F_{0}k_{\rm cz}/\eta_{\rm X},
\end{equation}
\begin{equation}\label{(20)}
     \tau_{{\rm{c,em}}}=t_{\rm b}/(1+z),
\end{equation}
where $\eta_{\rm X}\equiv \int_{0.3{\rm~{keV}}}^{10{\rm~{keV}}}L_{\nu}d\nu/L_{{\rm{EM}}}$ is the radiative efficiency of the ejecta in the XRT band \citep{2014MNRAS.443.1779R} \footnote{Here, the radiative efficiency $\eta_{\rm X}$ is very difficult to constrain, so that we adopt $\eta_{\rm X}=10^{-2}$ in our calculations \citep{2019ApJ...878...62X}.} and $D_{\rm L}$ is the luminosity distance of the burst.

\begin{figure*}
    \vspace{2cm}
	\subfigbottomskip=1pt
	\subfigcapskip=-5pt
	\centering
\includegraphics[height=7cm,width=18cm]{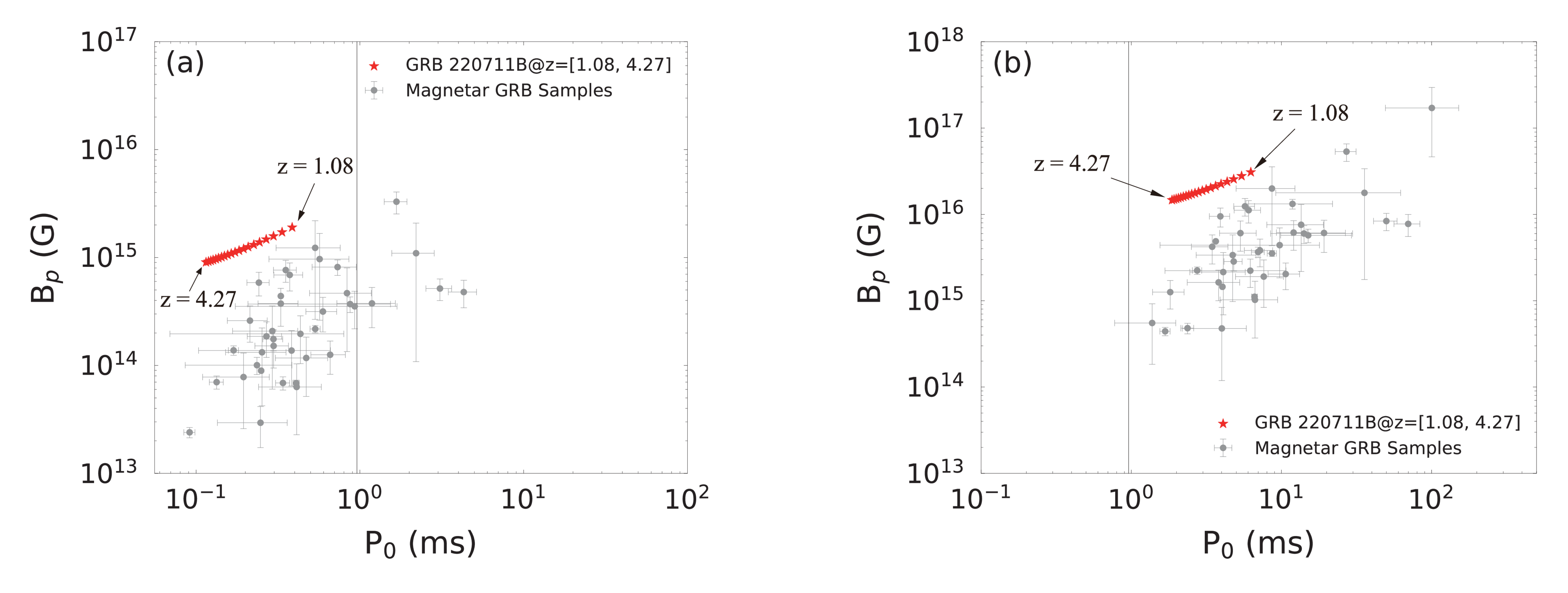}
\caption{Initial spin period $P_{{\rm{0}}}$ versus surface polar cap magnetic field strength $B_{{\rm{p}}}$ of GRB 220711B and magnetar samples taken from \protect\cite{2014ApJ...785...74L}. The left and the right panels correspond to isotropic case and beaming corrections with jet opening angle $5^{\circ}$, respectively. The vertical black lines represent the breakup spin-period for a magnetar.}\label{figure:6} 
\end{figure*}

Combining with Eqs. (\ref{(15)}) and (\ref{(16)}), we derive $P_{{\rm{0}}}$ and $B_{{\rm{p}}}$ for a series of redshifts ranging from 1.08 to 4.27 by adopting the typical values of the parameters \citep{2021MNRAS.506.5268R}, such as $M=1.4M_{\odot}$, $R=10^{6}{\rm{cm}}$, and $I\simeq \frac{2}{5}MR^2$. The derived results are shown in Table 2. Over the range in redshifts, we find that $P_{{\rm{0}}}$ and $B_{{\rm{p}}}$ lie in the range [0.12, 0.39] ms and [$9.05\times 10^{14}$, $1.90\times 10^{15}$] G, respectively. We compare the values of $P_{{\rm{0}}}$ and $B_{{\rm{p}}}$ of GRB 220711B with that of other typical long-duration GRBs derived by \cite{2014ApJ...785...74L}. We find that the spread in $P_{{\rm{0}}}$ values does not fall into the expected reasonable range (see Figure \ref{figure:6}a), and all of the simulated data are already exceeded the breakup spin-period for a neutron star \citep{2004Sci...304..536L}. Therefore we must consider beaming corrections for this case. 

Since we do not observe the jet break signature at the end of the afterglow, we choose $\theta_{\rm j}\sim5^{\circ}$ which is a typical jet opening angle of most long-duration GRBs to account for the beaming corrections \citep{2001ApJ...562L..55F,2008ApJ...675..528L}. The beaming factor $f_{\rm b}$ can be roughly estimated as $f_{\rm b}=1-\rm{cos}\theta_{\rm j}\simeq \frac{1}{2}\theta_{\rm j}^{2}$, and Eq. (\ref{(17)}) is replaced by $L_{\rm X,corr}(t)=L_{\rm X}(t)f_{\rm b}=4\pi D_{\rm L}^{2}F_{{\rm{0}}}f_{\rm b}k_{\rm cz}$. Following the same method above, we calculated $P_{{\rm{0}}}$ and $B_{{\rm{p}}}$ again by considering the beaming correction. We find that $P_{{\rm{0}}}$ and $B_{{\rm{p}}}$ lie in the range [1.87, 6.25] ms and [$1.47\times 10^{16}$, $3.09\times 10^{16}$] G, respectively within the redshift range [1.08, 4.27]. The parameter of $B_{{\rm{p}}}$ is slightly larger than that of other typical long-duration GRBs, while $P_{{\rm{0}}}$ fall in a reasonable range (Figure \ref{figure:6}b).

At 500 seconds after the trigger time, the X-ray emission continues to decay with a steeper index $\sim t^{-11}$ followed by shallow power-law decay with slope index $\sim t^{-0.9}$. The steeper decay with $\sim t^{-11}$ may be the signature of a supramassive magnetar collapsing into the black hole when the magnetar central engine can no longer prevent gravitational collapse \citep{2007ApJ...665..599T,2010MNRAS.409..531R,2013ApJ...763L..22Z,2017ApJ...849...71L}. The final segment of X-ray emission with a shallow decay $\sim t^{-0.9}$ is consistent with the external shock model when the relativistic ejecta of the jet is decelerated by the circumburst medium \citep{1997ApJ...476..232M,1998ApJ...497L..17S}.

\subsection{Precession motion of a magnetar with QPO}\label{3.3}
The features of the X-ray afterglow emission suggest that the supermassive magnetar seems to survive for $\sim$500s as the central engine of GRB 220711B, and both $\gamma-$ray and X-ray light curves show a significant QPO signature with a high confidence level. Within this scenario, one basic question is how can such a QPO signal be produced by the magnetar central engine?

A natural interpretation of such a QPO signal invokes a precession motion of the magnetar central engine, which has been proposed and studied in many publications \citep{1970ApJ...160L..11G,2012ApJ...760...63L,2015MNRAS.451..695Z,2020ApJ...892L..34S,2022MNRAS.513L..89Z}. If a newborn magnetar is spinning rapidly and is initially non-spherical, highly magnetized, and has a large ellipticity, then the inclination angle $\phi$ between the rotation axis, \textbf{$\Omega_{R}$}, and magnetic axis \textbf{B} gives rise to an oscillation in the GRB $\gamma$-ray or X-ray flux,
\begin{equation}\label{(19)}
L_{{\rm{EM}}}(t)=\eta_{X}\frac{B_{{\rm{p}}}^{2}R^{6}\Omega_{0}^{4}}{6c^{3}}\,(1+\frac{t}{\tau_{{\rm{c,em}}}})^{-\alpha}\,\lambda(\delta, \phi_{0}, k, \Omega_{p}),
\end{equation}
where $\alpha$ is the slope index after the break, and
\begin{equation}\label{(20)}
\begin{aligned}
    \lambda(\delta, \phi_{0}, k, \Omega_{p})&=1+\delta\,\rm{sin}^{2}\phi\\
    &\approx 1+\delta[1-(\rm{cos}\phi_{0}+\textit{k}(\rm{cos}(\Omega_{p}t)-1))^{2}] 
\end{aligned}
\end{equation}
 is a factor dependent on inclination angle $\phi$ and quantifies the charge of the magnetosphere through the parameter $\delta$ ($\left|\delta\right|\leq 1$) \citep{2014MNRAS.441.1879P,2015MNRAS.453.3540A,2020ApJ...892L..34S,2022MNRAS.513L..89Z}.
$\phi_{0}$ is the initial inclination angle, $k$ is an order-unity factor related to the Euler angles \citep{2015MNRAS.451..695Z}, and $\Omega_{p}$ is the angular frequency of precession. It is worth noting that the angular frequency of precession is also possible time dependent \citep{2020ApJ...892L..34S}. On the other hand, the periods on the order of seconds have also been observed in some GRBs \citep{1991PhR...206..327H,1992Natur.357..472U,2008AIPC.1065..259M}. 

\begin{figure*}
    \vspace{1cm}
	\subfigbottomskip=5pt
	\subfigcapskip=-5pt
	\centering
\includegraphics[height=10cm,width=12.5cm]{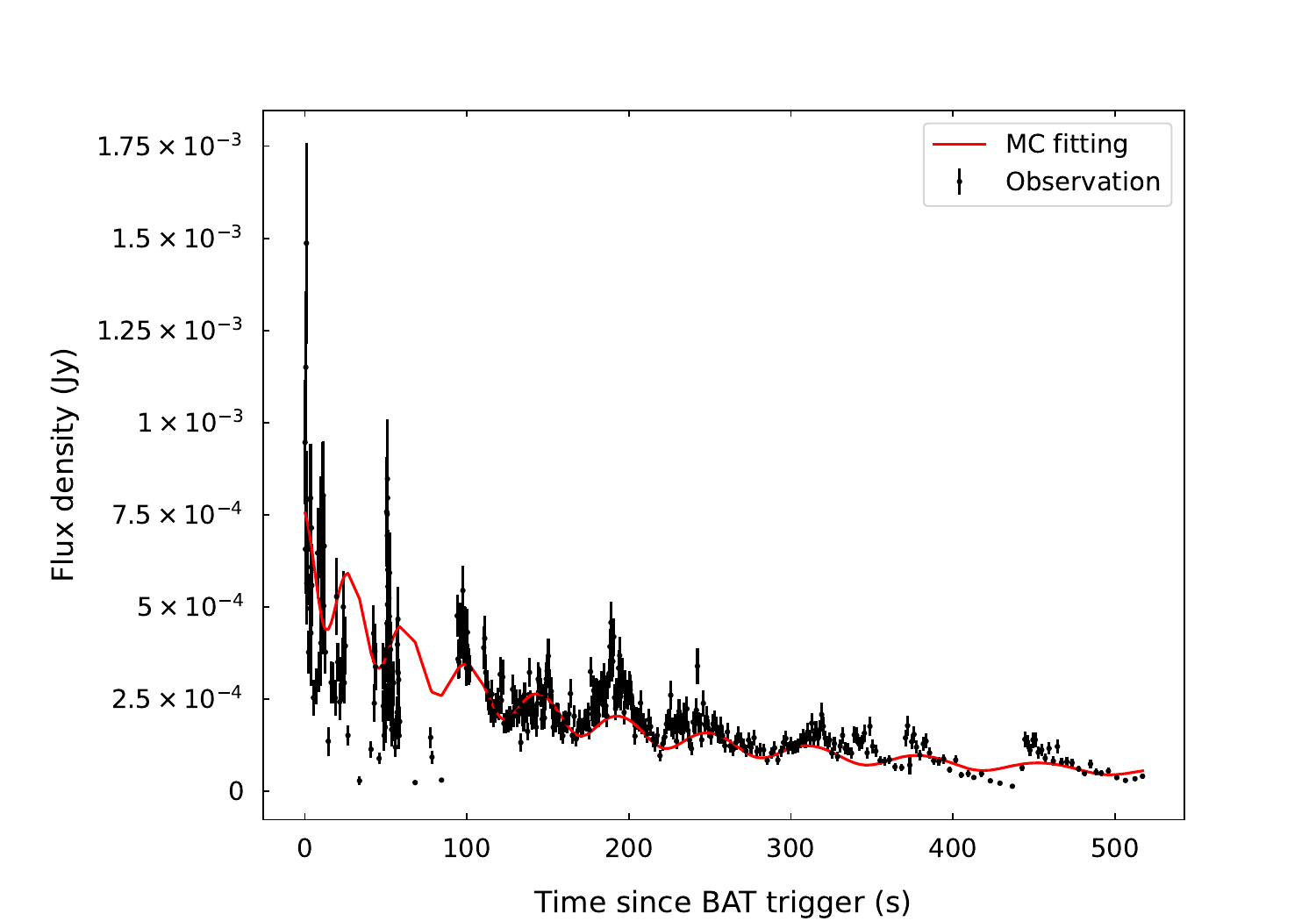}
\caption{MC fitting result with the magnetar precession model for BAT+XRT data.}\label{figure:7} 
\end{figure*}

Following a similar method from \cite{2022MNRAS.513L..89Z}, we adopt the {\sc Python} package {\sc EMCEE} \citep{2013PASP..125..306F} based on the Monte Carlo (MC) algorithm to fit the data. By considering the precession period being time dependent, namely, $\Omega_{\rm p} (t)\approx \epsilon \Omega_0 (1+t/t_{\rm c,em})^{1/2}$ \citep{2020ApJ...892L..34S}, where $\epsilon \ll 1$ and $\Omega_0=2\pi/P_{\rm{0}}$ are the ellipticity and initial spinning angular velocity of the magnetar, respectively. We fixed $R$, $I$, and $\eta_{X}$ to be $10^{6}\,{\rm{cm}}$, $1.11\times 10^{45}\,{\rm{g\cdot cm^{2}}}$, and $0.01$, respectively. By assuming a redshift of $z=1.08$, we obtain the following  magnetar parameters: $B_{{\rm{p}}}\sim 3.2\times 10^{16}\,{\rm{G}}$, $P_{{\rm{0}}}\sim 1.4\, {\rm{ms}}$, $\phi_{0}\sim 11.46^{\circ}$, $\delta\sim 0.89$, $k\sim 0.20$, and $\epsilon\sim 5.02\times 10^{-5}$. The values of $B_{{\rm{p}}}$ and $P_{{\rm{0}}}$ are roughly consistent with the range we derived in section \ref{3.2}. The fitting result is shown in Figure \ref{figure:7}.  

\section{Discussion and Conclusions}\label{4}
\subsection{Discussion}\label{4.1}
\cite{2024ApJ...970....6X} claimed that a QPO signal was discovered in the precursor of GRB 211211A which is believed to originate in a compact star merger \citep{2022Natur.612..232Y,2023ApJ...943..146C}. They proposed that the QPO signal comes from a catastrophic flare accompanied with torsional or crustal oscillations of the magnetar, and that this QPO signal should be only produced before the main GRB emission. The observed QPO signal in the X-ray emission of GRB 220711B peaking at $\sim 50$ s seems to be inconsistent with the above model. 

Furthermore, several physical models have been proposed to explain QPOs in magnetars, such as the magnetospheric plasma oscillation model \citep{1995MNRAS.275..255T}, matter-ring oscillation model \citep{1999PhR...311..259W}, magnetospheric Alfven wave oscillation model \citep{2021Natur.600..621C}, and the rotation modulation model \citep{1998ApJ...508..791M}. However, those models have trouble explaining the observed QPOs in GRBs. For example, the magnetospheric oscillation and matter-ring oscillation models require a stable magnetosphere or matter-ring to generate the oscillations. It is difficult to guarantee their existence in GRB prompt emission because the prompt emission can disrupt the magnetosphere or matter-ring due to its intense radiation. Both magnetospheric Alfven wave oscillation and rotation modulation models predict that observed QPOs have a high frequency up to kilohertz. This is inconsistent with the low frequency QPOs in GRB 220711B.

On the other hand, binary systems, such as a black hole with a stellar companion have also been proposed to produce QPOs in the X-ray emission from a few mHz to a few tens of kHz  \citep{2022MNRAS.510..807S}. The QPOs produced in this model are comparable with that of the observed QPO signal peaking at $\sim 50$ s, however, the central engine of such a binary system must be a Kerr black hole. This is inconsistent with GRB 220711B since we find significant evidence for it being powered by a magnetar central engine. Thus, a black hole with a stellar companion in a binary system should also be ruled out as the mechanism producing both a QPO signal and the extraordinary X-ray emission in GRB 220711B. 

Models of episodic accretion onto the compact stars (or young stellar objects) have also been proposed for interpreting the flares and QPOs in the X-ray emission (or in infrared and optical wavelengths) \citep{2010MNRAS.406.1208D,2013ApJ...768...63L,2014ApJ...789..129C}. However, the typical timescale of such episodic accretion onto compact stars (or young stellar objects) is of an order of one second (or hundreds of years), which is inconsistent with the observed QPO at $~50$ s. Thus, the episodic accretion model for interpreting the QPO in GRB 220711B can be ruled out.

\subsection{Conclusions}\label{4.2}
GRB 220711B is a long-duration GRB with a duration of $\sim$105 seconds, observed by both \textit{Fermi}/GBM and \textit{Swift}/BAT without a redshift measurement. We presented a broadband analysis of its prompt and afterglow emission, and found that the peak energy of its spectrum in the prompt emission is as low as $E_{\rm{p}}=79\,$ keV, which is softer than most long-duration GRBs observed by \textit{Fermi}/GBM \citep{2020ApJ...898L...6L}. By analysing the light curve of prompt emission which is characterized by a multi-peaked structure, no high-confidence-level QPO signal is found in the light curves of the prompt emission (e.g., BAT and GBM). More interestingly, the X-ray afterglow light curve is composed of a plateau emission smoothly connected with a $\sim t^{-2}$ segment overlapping some flares and followed by an extremely steep decay. It is found that a QPO signal at $\sim$ 50 s above 6$\sigma$ confidence level indeed exist in the X-ray afterglow. The temporal feature of X-ray light curve, together with the high-confidence-level QPO signal, is consistent with the prediction of a precessing magnetar central engine.

However, a fly in the ointment is that no redshift was measured in this case, so we have to use a pseudo redshift to reveal its physical properties. Several main results are summarized as follows:
\begin{itemize}	
\item[$\bullet$] Based on the features of the X-ray emission, we propose that the central engine of GRB 220711B is supramassive magnetar which can survive several hundred seconds. The plateau emission followed by a $\sim t^{-2}$ decay phase in the X-ray emission is consistent with the magnetar losing rotational energy via EM radiation, and the magnetar collapsing into the black hole at about 500 s corresponds to the abrupt drop decay. The final power-law decay segment of the X-ray emission is consistent with the external shock model. 
\item[$\bullet$] Based on the QPO signal in the early X-ray emission, we suggest that the magnetar central engine of GRB 220711B is precessing. 
\item[$\bullet$]We assume that the pseudo redshift lies in the range $1.08\leq z\leq4.27$, which is the 1$\sigma$ range of the redshift distribution of the \textit{Swift} long GRB samples. By considering the k-correction, X-ray radiative efficiency, and beaming correction, one can roughly estimate $P_{{\rm{0}}}$ and $B_{{\rm{p}}}$ of the magnetar lie in the range [1.87, 6.25] ms and [$1.47\times 10^{16}$, $3.09\times 10^{16}$] G, respectively. The parameter of $B_{{\rm{p}}}$ is slightly larger than that of other typical long-duration GRBs, but $P_{{\rm{0}}}$ fall in a reasonable range.
\end{itemize}

Moreover, the magnetar remains as a potential source of high-frequency, weak and continuous gravitational-wave when the magnetar is spinning fast \citep{2020ApJ...898L...6L}. This potential signal will play a critical role in understanding the physics of neutron stars. \cite{2022PhRvD.105j3019H} presented more details for calculating the GW radiation from a magnetar by considering the magnetically induced deformation, starquake-induced ellipticity, and accretion column-induced deformation. Any GW radiation from a magnetar has still not been detected by the current aLIGO and Virgo detectors, but is expected to be detected by the next generation of more sensitive GW detectors, such as the Einstein Telescope.

Upon finishing review of this paper, we were drawn attention to \cite{2025ApJ...985...33G}, who performed an independent analysis to search for QPO in the X-ray emission of GRB 220711B, and also found a QPO signal peaking at $\sim 50$ s in the early X-ray emission. They suggested that a stellar-merger-induced core collapse is a possible formation channel to produce the QPO in GRB 220711B.

\section*{Acknowledgements}
We acknowledge the use of the public data from the \textit{Swift} and \textit{Fermi} data archive, and the UK \textit{Swift} Science Data Center. We thank RuiChong Hu, JiGui Cheng, GuoYu Li, TianCi Zheng, WenYuan Yu, ZhaoWei Du, and WenHao Chen for helpful discussions. This work is supported by the Guangxi Science Foundation (grant No. 2023GXNSFDA026007), the Natural Science Foundation of China (grant Nos. 12494574, 11922301 and 12133003), the Program of Bagui Scholars Program (LHJ), and the Guangxi Talent Program (“Highland of Innovation Talents”).

\section*{Data Availability}
The data what we adopt is publicly available data from \textit{Fermi}/GBM, \textit{Swift}/BAT, and \textit{Swift}/XRT. There are no new data associated with this article. If one needs to adopt the data in this article, it should
be to cite this reference paper.

\bibliographystyle{aasjournal}

\clearpage

\begin{table*} 
   \begin{center}
			\caption{Time-integral spectral fitting results of GRB 220711B}
             \label{tab:1}
    \end{center}
 \begin{tabular}{cccccccc}
 		\hline
		\hline
 	$t_{1}\sim t_{2}$&Model&$\alpha_{0}$&$\beta_0$&$E_{p}$&kT&$\Delta$BIC& Favorite Model\\(s)&&&&(keV)&(keV)&&\\
   		\hline
		\hline
    $-40\sim 60$ &Band &$-0.57\pm0.5$ &$-2.34\pm0.61$       &$35^{+24}_{-24}$  &...  &429    &   \\
    &PL  &$-1.73\pm0.03$&... &...  &...   &507 &  \\
    &CPL &$0.78\pm0.14$&... &$75\pm11$ &... &423        &$\checkmark$     \\
    &BB &...&... &...  &$20.15\pm0.11$   &535  &  \\     
    &Band+BB &$1.20\pm0.29$  &$-6.02\pm2.46$ &...  &$32.65\pm2.98$ &427  &   \\        &PL+BB &$1.86\pm0.07$ &...    &... &$20.99\pm0.72$ &449   &     \\
    &CPL+BB &$0.56\pm0.22$  &...  &$55\pm20$    &$209.66\pm170.95$  &429&        \\
        \hline
 \end{tabular}
 \end{table*}

 \begin{table*} 
  \centering
  \caption{The derived Magnetar parameters of GRB 220711B for given different redshifts}
  \label{tab:magnetar_params}
  \begin{tabular}{cccccccc}
    \hline
    \hline
    Redshift ($z$) & $k_{\mathrm{cz}}$ & $L_{\mathrm{em,0}}$ & $\tau_{\mathrm{c,em}}$ & $B_{\mathrm{p}}$ & $P_{0}$ & Corrected $B_{\mathrm{p}}$ & Corrected $P_{0}$ \\
    & & ($10^{51}$ erg s$^{-1}$) & (s) & ($10^{15}$ G) & (ms) & ($10^{15}$ G) & (ms) \\
    \hline
    \hline
    1.080 & 0.923 & 1.57 & 94 & 1.90 & 0.39 & 30.86 & 6.25 \\
    1.250 & 0.915 & 2.26 & 87 & 1.72 & 0.34 & 27.85 & 5.43 \\
    1.420 & 0.907 & 3.10 & 81 & 1.58 & 0.30 & 25.59 & 4.81 \\
    1.580 & 0.901 & 4.03 & 76 & 1.48 & 0.27 & 23.91 & 4.35 \\
    1.750 & 0.895 & 5.18 & 71 & 1.39 & 0.24 & 22.48 & 3.96 \\
    1.920 & 0.889 & 6.51 & 67 & 1.31 & 0.23 & 21.30 & 3.64 \\
    2.090 & 0.883 & 8.01 & 63 & 1.25 & 0.21 & 20.31 & 3.38 \\
    2.260 & 0.878 & 9.70 & 60 & 1.20 & 0.20 & 19.48 & 3.15 \\
    2.420 & 0.873 & 11.45 & 57 & 1.16 & 0.18 & 18.81 & 2.97 \\
    2.590 & 0.869 & 13.50 & 54 & 1.12 & 0.17 & 18.18 & 2.81 \\
    2.760 & 0.864 & 15.75 & 52 & 1.09 & 0.16 & 17.63 & 2.66 \\
    2.930 & 0.860 & 18.19 & 50 & 1.06 & 0.16 & 17.15 & 2.53 \\
    3.095 & 0.856 & 20.75 & 48 & 1.03 & 0.15 & 16.73 & 2.42 \\
    3.260 & 0.853 & 23.51 & 46 & 1.01 & 0.14 & 16.35 & 2.32 \\
    3.430 & 0.849 & 26.55 & 44 & 0.99 & 0.14 & 16.00 & 2.22 \\
    3.600 & 0.845 & 29.80 & 42 & 0.97 & 0.13 & 15.68 & 2.14 \\
    3.770 & 0.842 & 33.26 & 41 & 0.95 & 0.12 & 15.39 & 2.06 \\
    3.934 & 0.839 & 36.80 & 40 & 0.93 & 0.12 & 15.13 & 1.99 \\
    4.100 & 0.836 & 40.60 & 38 & 0.92 & 0.12 & 14.89 & 1.93 \\
    4.270 & 0.833 & 44.70 & 37 & 0.91 & 0.12 & 14.67 & 1.87 \\
    \hline
  \end{tabular}
\end{table*}
%%%%%%%%%%%%%%%%%%%%%%%%%%%%%%%%%%%%%%%%%%%%%%%%%%%%%%%%%%%%%%%%%%%%%%%
%%%%%%%%%%%%%%%%%%%%%%%%%%Figures%%%%%%%%%%%%%%%%%%%%%%%%%%%%%%%%%%%%%%%%%%%%%
%%%%%%%%%%%%%%%%%%%%%%%%%%%%%%%%%%%%%%%%%%%%%%%%%%%%%%%%%%%%%%%%%%%%%%%%%%%%%%%%%%%%%%%%%%%%%%%%%%%%%%%%%%%%%%%%%%%%%%%%%%%%%%%%%%%%
%%%%%%%%%%%%%%%%%%%%%%%%%%%%%%%%%%%%%%%%%%%%%%%%%%%%%%%%%%%%%%%%%%
%%%%%%%%%%%%%%%%%%%%%%%%%%%%%%%%%%%%%%%%%%%%%%%%%%%%%%%%%%%%%%%%%%
%%%%%%%%%%%%%%%%%%%%%%%%%%%%%%%%%%%%%%%%%%%%%%%%%%%%%%%%%%%%%%%%%%%%%%%%%%%%%%%%%%%%%%%%%%%%%%%%%%%%%%%%%%%%%%%%%%%%%%%%%%%%%%%%%%%%
%%%%%%%%%%%%%%%%%%%%%%%%%%%%%%%%%%%%%%%%%%%%%%%%%%%%%%%%%%%%%%%%%%%%%%%%%%%%%%%%%%%%%%%%%%%%%%%%%%%%%%%%%%%%%%%%%%%%%%%%%%%%%%%%%%%%

%%%%%%%%%%%%%%%%%%%%%%%%%%%%%%%%%%%%%%%%%%%%%%%%%%%%%%%%%%%%%%%%%%%%%%%%%%%%%%%%%%%%%%%%%%%%%%%%%%%%%%%%%%%%%%%%%%%%%%%%%%%%%%%%%%%%
%%%%%%%%%%%%%%%%%%%%%%%%%%%%%%%%%%%%%%%%%%%%%%%%%%%%%%%%%%%%%%%%%%
%%%%%%%%%%%%%%%%%%%%%%%%%%%%%%%%%%%%%%%%%%%%%%%%%%%%%%%%%%%%%%%%%%
%%%%%%%%%%%%%%%%%%%%%%%%%%%%%%%%%%%%%%%%%%%%%%%%%%%%%%%%%%%%%%%%%%%%%%%%%%%%%%%%%%%%%%%%%%%%%%%%%%%%%%%%%%%%%%%%%%%%%%%%%%%%%%%%%%%%
%%%%%%%%%%%%%%%%%%%%%%%%%%%%%%%%%%%%%%%%%%%%%%%%%%%%%%%%%%%%%%%%%%%%%%%%%%%%%%%%%%%%%%%%%%%%%%%%%%%%%%%%%%%%%%%%%%%%%%%%%%%%%%%%%%%%

\label{lastpage}
\end{document}